\documentclass[twocolumn,amsmath,amssymb,
aps,
pra,
10pt,showkeys,superscriptaddress,longbibliography
,nofootinbib
,floatfix
]{revtex4-2}
\usepackage{dcolumn}
\usepackage{bm}
\usepackage{mathtools, cases}
\usepackage[unicode]{hyperref}
\usepackage{graphicx}
\usepackage{soul}
\hypersetup{colorlinks=true,linkcolor=blue,citecolor=blue,filecolor=black,urlcolor=blue}
\renewcommand{\d}{\mathrm{d}}

\begin{document}

\title{The influence of packing protocol, size ratio, and pore structure on fractal exponents in dense polydisperse packings}

\author{Artem A. Vladimirov}
\affiliation{Joint Institute for Nuclear Research, 141980 Dubna, Russian Federation}
\affiliation{Dokuchaev Soil Science Institute, 119017 Moscow, Russian Federation}

\author{Alexander Yu. Cherny}
\email[Corresponding author, e-mail:~]{cherny@theor.jinr.ru}
\affiliation{Joint Institute for Nuclear Research, 141980 Dubna, Russian Federation}

\author{Eugen M. Anitas}
\affiliation{Joint Institute for Nuclear Research, 141980 Dubna, Russian Federation}
\affiliation{Horia Hulubei National Institute of Physics and Nuclear Engineering, 077125 Bucharest-Magurele, Romania}
\affiliation{Department of Physics, Craiova University, 200585 Craiova, Romania}

\author{Vladimir A. Osipov}
\affiliation{Joint Institute for Nuclear Research, 141980 Dubna, Russian Federation}

\date{\today}

\begin{abstract}
We study fractal properties of systems of densely and randomly packed disks, obeying a power-law distribution of radii, which is generated by using various protocols: Delaunay triangulation (DT) and constant pressure (CP) protocols and the generalized Apollonian packing. The power-law exponents of the mass-radius relation and structure factor are obtained numerically for various values of the size ratio of the distribution, defined as the largest-to-smallest radius ratio. We show that the size ratio is an important control parameter responsible for the consistency of the fractal properties of the system: the larger the ratio, the less pronounced the finite-size effects and the better the agreement between the exponents. For the DT protocol, all three exponents coincide even at moderate values of the size ratio. For the CP protocol, the exponents are different for both moderate and large size ratios. The suppression of the exponent of the structure factor in the CP packing is explained by the specific behaviour of pores, which contain relatively large cavities. We develop an algorithm for calculating the pore size distribution and show that it is related to the exponent of the structure factor. We argue that the presence of the cavities lowers the configurational entropy and thus reduces the randomness of the CP packing. Thus the cavities reduce both packing fraction and randomness of the CP packings. Nevertheless, there is a tendency for the exponents to converge as the size ratio increases, suggesting that all the exponents become equal in the limit of an infinite size ratio.
\end{abstract}


\maketitle

\section{Introduction}
\label{sec:intro}

Compact packings with power-law size distributions are of great interest for various applications in science and engineering, in particular for understanding the structural properties of materials \cite{torquato18}. This understanding is critical for applications ranging from ultra-high performance concrete \cite{qifeng21} to biological systems \cite{sato18} and soils \cite{Wang22}, as it enables the design and optimization of materials with desired mechanical properties and stability.

Analytical solutions were obtained for the jamming fractions of polydisperse packings of spheres \cite{Anzivino23} and disks \cite{Zaccone25} in the case of moderate variations of their sizes (see also Ref.~\cite{Farr09}). The case of a high size ratio of the power-law polydispersity was considered in the papers \cite{cherny22,cherny23,monti23,cherny24}. The spatial correlations in such systems were found to exhibit fractal-like behaviour. The description of the emerging fractal properties is one of the most interesting and intriguing problems. In this paper, we continue to explore these issues.

In numerical simulations, dense packing can be achieved by various methods (see \cite{torquato18} for a review), including
random successive addition (RSA) \cite{Renyi58,Widom66}. In RSA, particles are added one at a time in random positions. If a particle overlaps with existing ones, it is discarded. This process continues until no more particles can be added without overlap. Recently, the authors suggested \cite{cherny24} the Delaunay triangulation protocol as a modification of RSA. DT creates a network of triangles connecting the centers of packed particles, which optimizes the process of finding empty spaces for newly added particles.

Another method of dense packings is the constant pressure protocol, which uses LAMMPS (large-scale atomic/molecular massively parallel simulator) with the GRANULAR package \cite{Thompson22}. It allows the modeling of granular materials under controlled pressure conditions. Constant external pressure slowly compresses an initially dilute configuration of particles, leading to a compact jammed packing.

We also consider the generalized Apollonian packing (GAP) \cite{oron2000}, which  is a deterministic construction that extends the classical Apollonian packing \cite{mandelbrot82} and enables us to get the fractal exponents up to $1.803\ldots$. Unlike the DT and CP protocols, GAP does not pack a given set of disks, but generates it deterministically. The GAP is included in this paper mainly as a reference for pore size analysis (see Sec.~\ref{sec:pores} below) but not as a random packing protocol. We consider a construction of GAP, which yields the exponent $D=1.46$, quite close to $D=1.5$ (see Sec.~\ref{sec:GAP} below).

It was shown with RSA \cite{cherny23} and DT \cite{cherny24} protocols that \emph{dense random} packings of disks with a power-law size distribution exhibit fractal-like properties with the fractal dimension $D_{\mathrm{f}}$ being equal to the exponent of the power-law size distribution $D$. On the other hand, Monti et al. \cite{monti23} obtained jammed packings of the same distribution with the CP protocol and argued that the packings, although exhibiting fractal properties, nevertheless have fractal dimensions different from the exponent of the power-law size distribution. The authors of the paper \cite{monti23} concluded that the fractal exponent depends on packing protocol.

Fractal dimension $D_{\mathrm{f}}$ of a fractal is \emph{defined} through the exponent in the mass-radius relation \cite{mandelbrot82,gouyet96:book}: $M(r)\sim r^{D_{\mathrm{f}}}$, which coincides with the exponent $D_{\mathrm{f}}$ of decay of the structure factor $S(q)\sim 1/q^{D_{\mathrm{f}}}$. The relation between both exponents is based on Erd\'elyi's theorem for asymptotic expansion of Fourier integrals \cite{erdelyi56:book}, because the structure factor is the Fourier transform of the pair distribution function, which is proportional to $\frac{\partial M(r)}{\partial r}\frac{1}{r}$, see Appendix \ref{sec:gen_rel}.

In practice, however, the fractal power-law behaviour is always realized within a \emph{finite} fractal range, and thus the relation between the power-law exponent of the structure factor and fractal dimension should be used with some caveats: it is valid only if the fractal range in momentum space is \emph{sufficiently long}.  This ``rule of thumb'' is well known in experimental studies of fractal aggregates using small-angle scattering \cite{schmidt91}, which directly yields the structure factor of the system under study.

As shown in the previous paper \cite{cherny23}, the fractal range of densely packed disks or spheres with a power-law size distribution is determined by their largest-to-smallest size ratio, and this result is still valid in the thermodynamic limit \cite{cherny24}. In the paper \cite{monti23}, the ratio was taken in two dimensions from $100$ to $300$ for various exponents $D$ of the distribution\footnote{Note that in the paper \cite{monti23}, the notation $\beta=D+1$ is used.}. It is shown below in Sec.~\ref{sec:monti} that these ratios are not large enough to obtain the correct value of the fractal dimension of the system. To this end, the data of jammed packings with the CP protocol of Ref.~\cite{monti23} are reproduced for a specific value of the exponent $D=1.5$ and the size ratio $292$. We calculate both structure factor and mass-radius relation and obtain the discrepancy between the exponents $\alpha$ [$S(q)\sim 1/q^{\alpha}$] and $D_{\mathrm{f}}$ [$M(r)\sim r^{D_{\mathrm{f}}}$] in the corresponding fractal ranges:  $\alpha\simeq1.31$  and $D_{\mathrm{f}}\simeq1.419$, see Fig.~\ref{fig:DT_CP_300} and Table \ref{tab:exponents} below. This discrepancy is due to the insufficient size ratio.

Increasing the size ratio to $R/a = 1575$ and keeping the other parameters unchanged (see Sec.~\ref{sec:CPP1500} below), we obtain $\alpha\simeq1.41$ and ${D_{\mathrm{f}}}\simeq1.441$, which are much closer to each other, although these values are still different from $D=1.5$ predicted by our model of dense random packings.

To understand the finite-size effects in CP packing, the pore size distribution is studied. We have developed an algorithm that effectively represents pores as a set of disks (see Sec.~\ref{sec:pores} below). It is shown that the CP protocol generates relatively large cavities, which affect the fractal exponents and decrease their values. The pore size distribution is strongly influenced by the size ratio of the particle distribution. Random packings generated by the DT protocol satisfy $D_\mathrm{f} = \alpha = D$ once the size ratio is sufficiently large, whereas CP packings exhibit some deviations due to the formation of large cavities that reduce randomness and modify the effective pore-size distribution. Nevertheless, as the size ratio increases, the influence of the cavities diminishes, and the fractal exponents obtained with different protocols show a \emph{clear tendency to converge} to $D$ (see Table \ref{tab:exponents} below).

Thus, although different packing protocols lead to clearly different configurations, the apparent dependence of fractal exponents on the protocol is not fundamental. Instead, the decisive factor determining the fractal properties of dense polydisperse packings is the degree of configurational randomness, which is closely related to the pore statistics.

This paper is organized as follows. In Sec.~\ref{sec:disks_meth}, we explain in detail a set of disks to be packed and packing methods. In the next section, the GAP is considered and the parameters of its generation are described. The fractal properties of dense random packing with the DT protocol are studied for various values of the size ratio in Sec.~\ref{sec:DT}. In Sec.~\ref{sec:monti}, the results for packings generated using the CP protocol are presented. In the next section, we find the pore size distribution in various protocols and examine its effect on the fractal exponent of the structure factor. In Conclusions, we summarize the main results and outline prospects for future research.

\section{A set of disks to be packed and methods}
\label{sec:disks_meth}

\begin{figure*}[tb]
\centerline{\includegraphics[width=0.33\textwidth,clip=true]{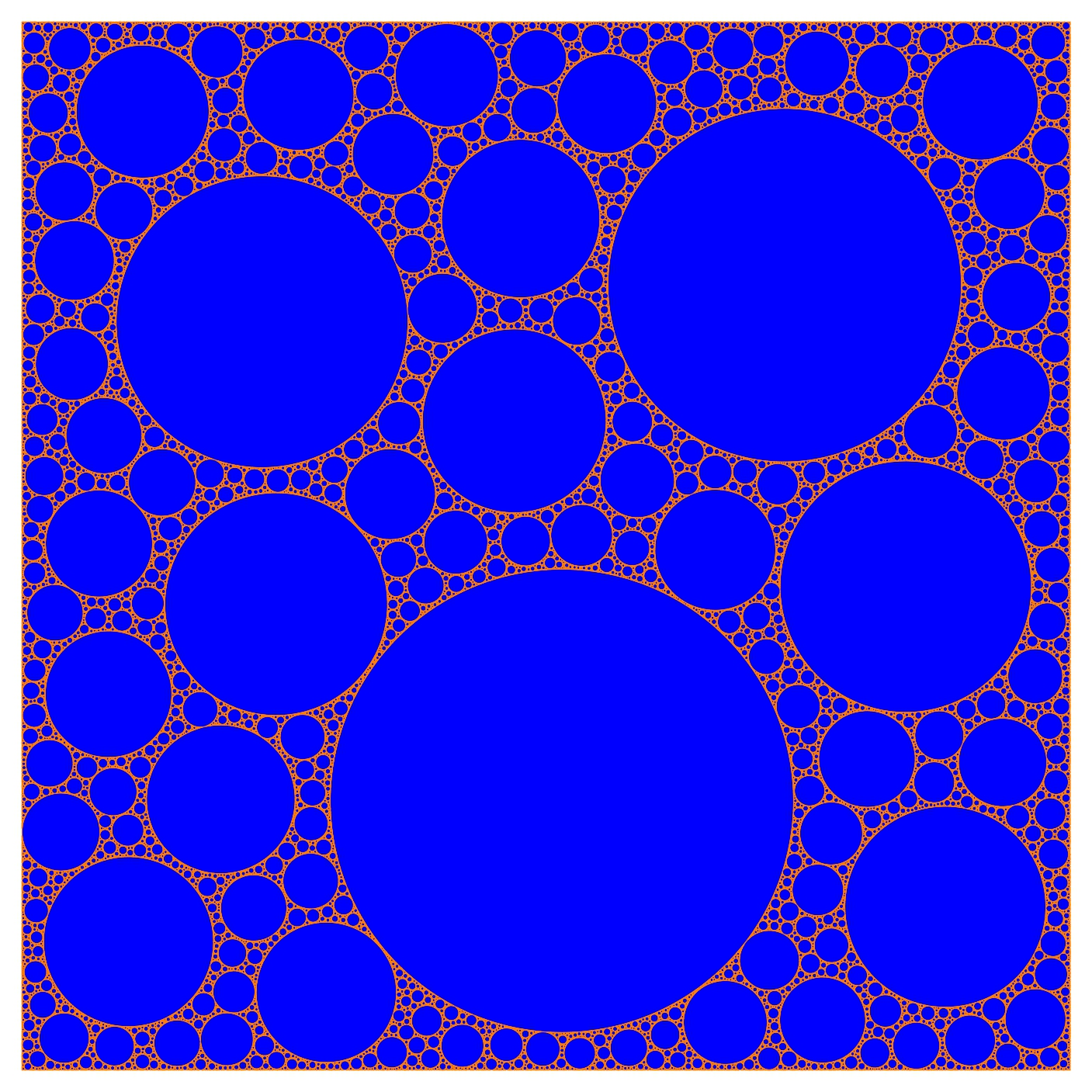}\includegraphics[width=0.33\textwidth,clip=true]{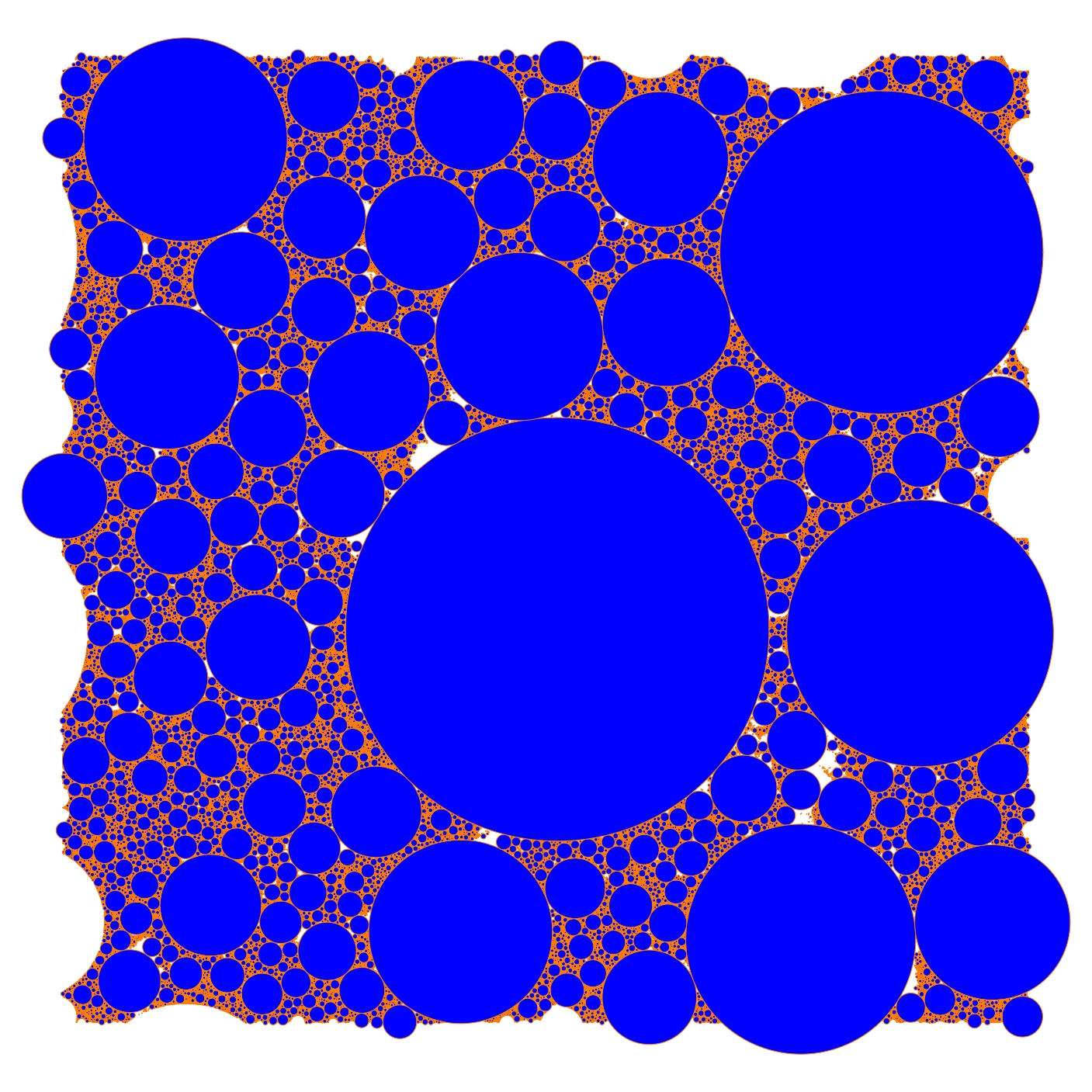}\includegraphics[width=0.33\textwidth,clip=true]{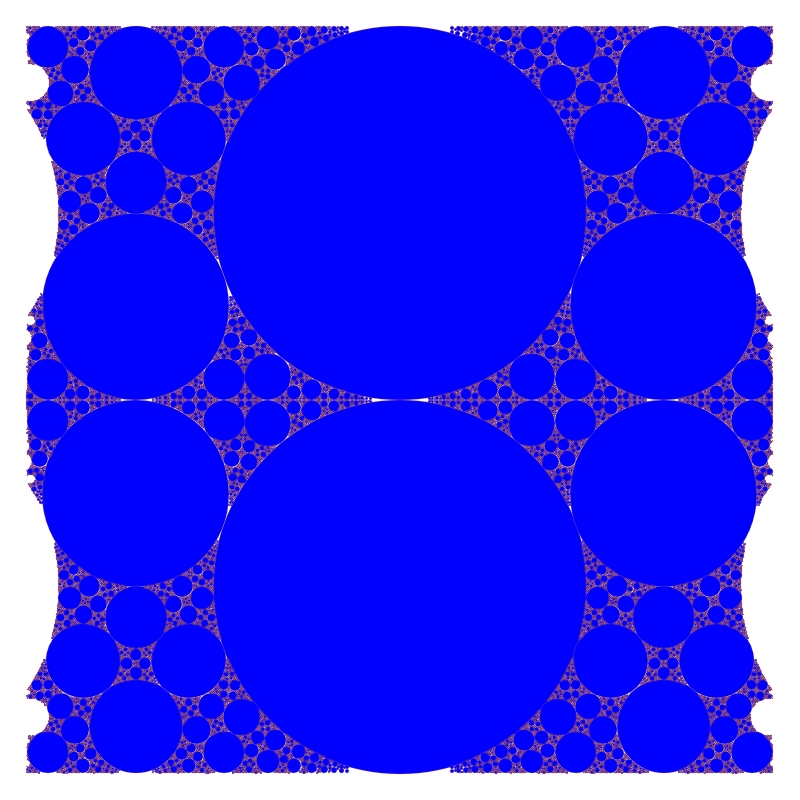}}
\caption{\label{fig:methods} Comparison of spatial distributions of $N = 125000$ disks obtained from different protocols: (a) DT with zero boundary conditions (packing fraction 0.983). (b) CP with periodic boundary conditions (packing fraction 0.967). The disk borders and pores are highlighted in orange and white, respectively. One can notice some inhomogeneities and voids in the configuration generated with the CP protocol, see the discussion in Sec.~\ref{sec:CPP300} below. (c) GAP (packing fraction 0.943) and N = 52366. In all cases the ratio between the largest and smallest radii is 1575.}
\end{figure*}

Let us consider a set of $N$ disks obeying a power-law distribution with the exponent $D$. We introduce a cumulative distribution and denote the number of disks, whose radii are bigger than or equal to $r$, by $N(r)$. Then we have $N(r)\sim 1/r^D$. In two dimensions, the exponent satisfies the condition $0<D<2$. The number of disks $|d N(r)|$ whose radii fall within the range ($r,r + d r$) is proportional to $dr/r^{D+1}$. The radii vary within the range of the distribution, from $a$ to $R$. The size ratio $R/a$ determines the length of the range of the distribution on a logarithmic scale. For a finite number of disks, a choice of specific radii that follow the power-law distribution was described in detail in Refs.~\cite{cherny23,cherny24}.

{For the RSA and DT protocols to enable compact packing within a finite area,} the exponent of the size distribution should be restricted to $D_{\mathrm{Ap}}< D < 2$ with $D_\mathrm{Ap}=1.3057\ldots$ being the dimension of the Apollonian packing \cite{Aste96} (see the discussion in Refs.~\cite{cherny23}).
The disks are arranged into a square by placing them from the largest to the smallest. As the set of already placed disks grows, checking their collisions takes more and more computational time. The DT algorithm reduces computation costs of searching for an empty space to add a new disk, which makes it especially effective at high packing fractions. The total number of operations for dense packing with and without DT is $O(N\log N)$ and $O(N^{2})$, respectively (see details of the DT protocol in our paper \cite{cherny24}). We employ the code from the CGAL software package \cite{cgal}.

For the constant pressure protocol, we employ the LAMMPS GRANULAR package following the procedure described in Ref.~\cite{monti23}. For particle interactions, the ``gran/hooke'' potential \cite{Silbert2001} is used to simulate frictionless, damped, and purely repulsive Hookean springs.  The normal force $F_\mathrm{n}$ between particles $i$ and $j$ of radii $R_i$ and $R_j$, respectively, separated by distance $r_{ij}$ is $F_\mathrm{n} = k_\mathrm{n} (R_i + R_j - r_{ij}) - M_{\mathrm{eff}}\gamma_\mathrm{n} v_\mathrm{n}$, where $k_\mathrm{n}$, $M_{\mathrm{eff}} = M_i M_j /(M_i + M_j )$, and $\gamma_\mathrm{n}$ are the spring stiffness, effective reduced mass, and damping coefficient, respectively.  The force is supposed to be purely repulsive, so $F_\mathrm{n} = 0$ if $k_\mathrm{n} (R_i + R_j - r_{ij}) - M_{\mathrm{eff}}\,\gamma_\mathrm{n} v_\mathrm{n}\leqslant 0$.  To apply external isotropic pressure $p_\mathrm{a}$ on the system, we use ``press/berendsen" barostat \cite{Berendsen1984}, which computes the internal pressure of the system and rescales the system volume and particle coordinates until the internal pressure matches the applied pressure. Packing stops when the kinetic energy per particle becomes small. The periodic boundary conditions are imposed on the system of disks.

We choose the following simulation parameters: $p_\mathrm{a} = 10^{-4} k_\mathrm{n} / a$, where $a$ is the smallest radius of the disks,  $\gamma_\mathrm{n}=0.5$, and the particle mass densities and spring stiffness are set to one, so that $M_i = \pi R_i^2$ and $k_\mathrm{n} = 1$.  The simulation box is a periodic square, initially filled with a dilute system of the disks, which are placed randomly without overlaps. We use the DT protocol to generate the initial spatial distribution of the disks. The initial packing fraction is chosen to be $\sim 0.66$, which is sufficient for the CP packing. Note that further reduction of the initial density increases the computation time but does not affect the characteristics of the resulting CP packings. Configurations obtained using the DT and CP protocols are shown in Fig.~\ref{fig:methods}.

\begin{figure}[tb]
\centerline{\includegraphics[width=.95\columnwidth,clip=true]{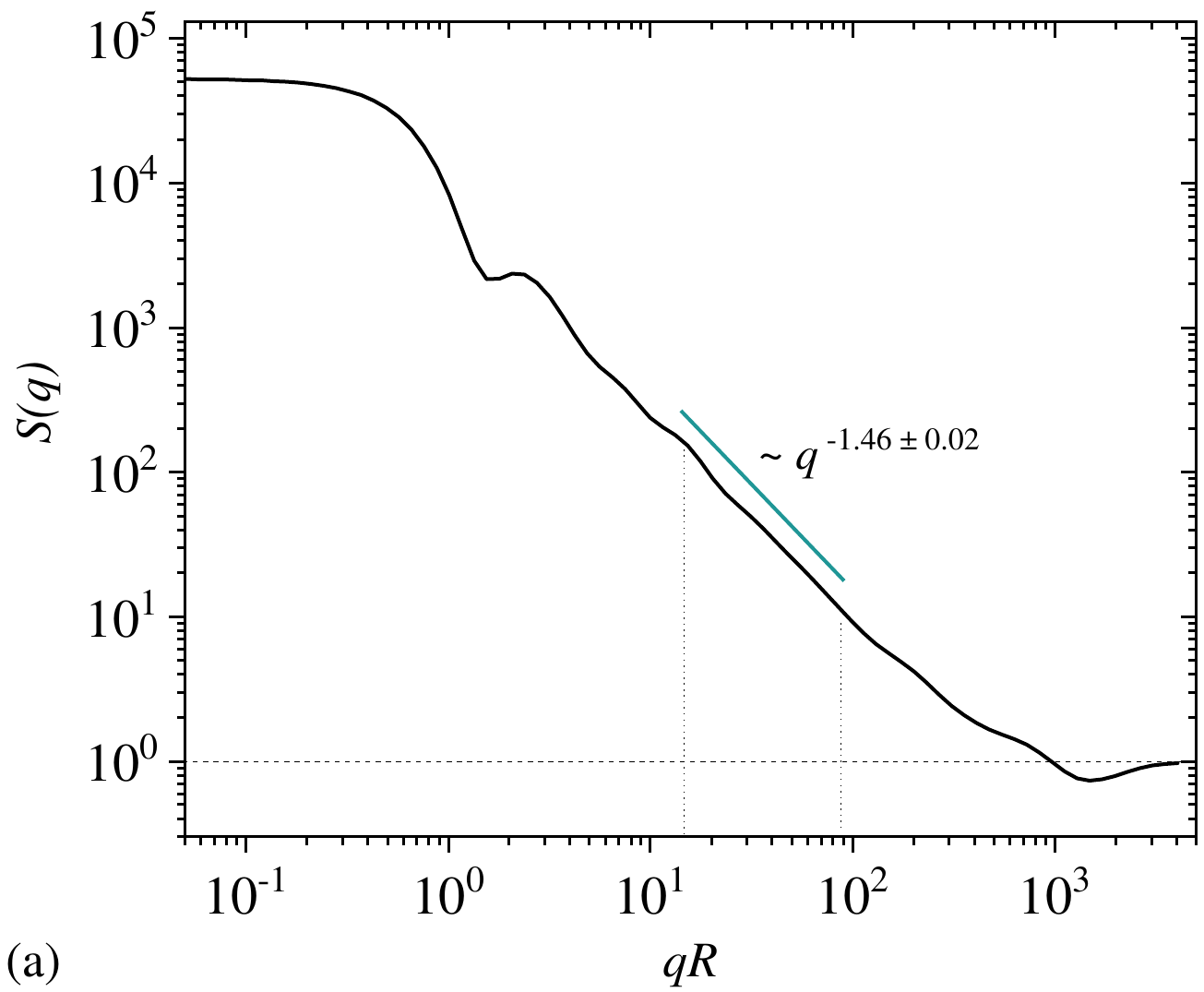}}
\centerline{\includegraphics[width=.95\columnwidth,clip=true]{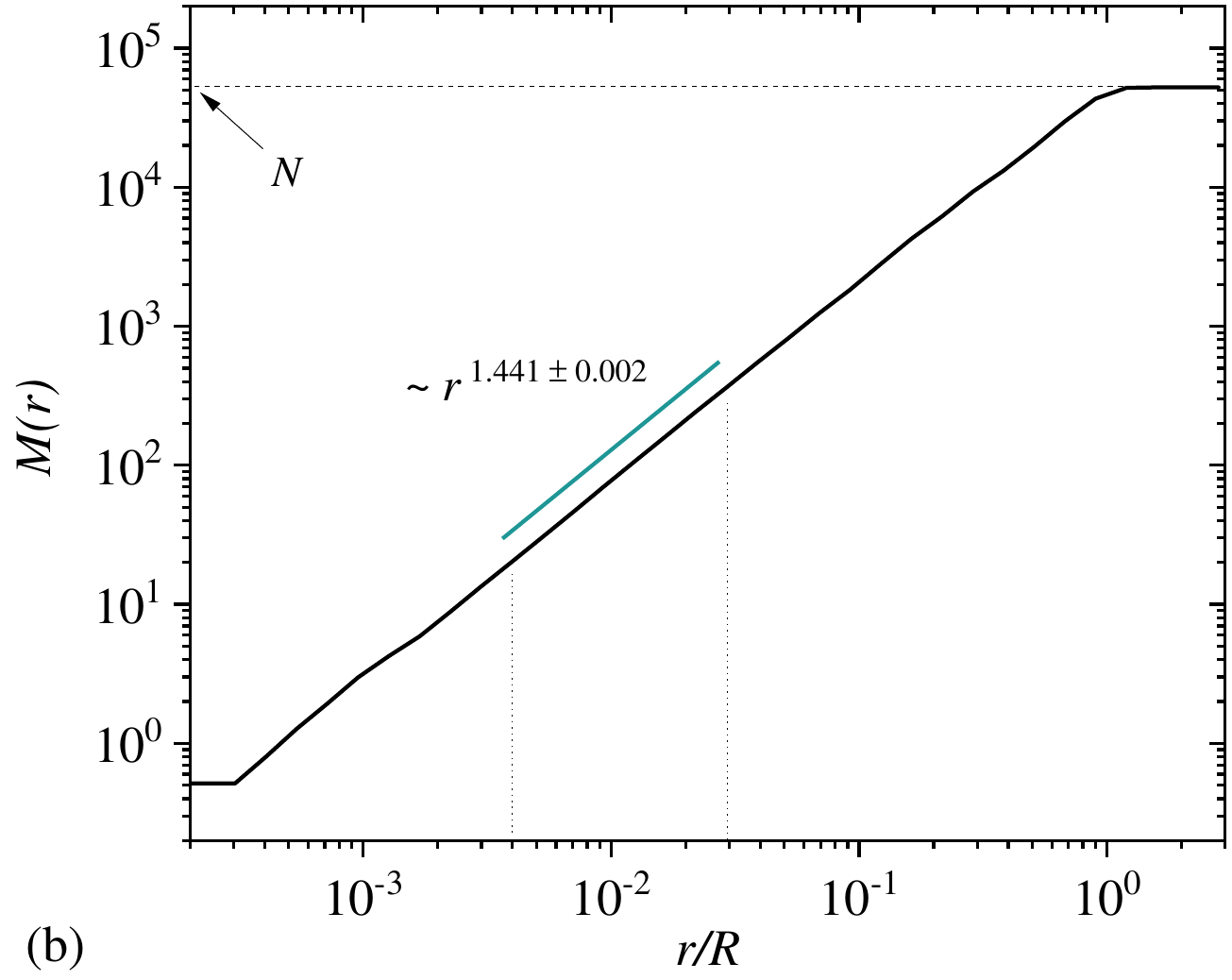}}
\caption{\label{figA2} The smoothed structure factor (a) and mass-radius relation (b) for GAP when the ratio between the largest and smallest radii is 1575. Vertical dotted lines denote the borders of the range over which the fit has been performed.}
\end{figure}

\section{Generalized Apollonian Packing}

\label{sec:GAP}

In addition to the random packing protocols discussed above, we also employ a deterministic hierarchical construction based on the generalized Apollonian packing, also known as a space-filling bearing. This deterministic structure provides a useful benchmark configuration, against which the statistical properties of the random packings generated by the DT and CP protocols can be compared.

We follow the algorithm of Oron and Herrmann \cite{oron2000}, which enables us to construct the sets with the fractal dimension ranging approximately from $1.306$ to $1.803$. The method generates a strip geometry that is invariant under a prescribed set of M\"obius transformations. The transformations are reflections and inversions chosen such that the resulting configuration contains no frustrated loops and completely fills the strip in the limit of infinite number of iterations. Below we describe the parameters of the algorithm that we used and refer the reader to the paper \cite{oron2000} for details.

The construction begins from a ``base loop'' consisting of $l$ circles, which are mapped onto themselves under two inversions $I_{1}$ and $I_{2}$ centered at the outermost contact points of the seed. For the F2 family used in this work, the seed is symmetric with respect to two vertical reflection axes, and the alignment of the contact points ensures that the inversion-circle centers lie on a line orthogonal to the strip boundaries. The parameters $n_{1}$ and $n_{2}$ determine the radii of the inversion circles through the relations $r_{1}^{2}=a^{2}z_{n_{1}}$ and $r_{2}^{2}=a^{2}z_{n_{2}}$, with $z_{k}=\left[\cos\!\left(\pi/(k+3)\right)\right]^{-2}$, and $a = (z_{n_1} + z_{n_2})^{-1/2}$ [see Eq.~(A.8) in Ref.~\cite{oron2000}] denoting half of the strip period. Repeated application of the transformation sequences $R_{1}I_{1}, (R_{1}I_{1})^{2}, \cdots, (R_{1}I_{1})^{n_{1}+2}$ and $R_{2}I_{2}, (R_{2}I_{2})^{2}, \cdots, (R_{2}I_{2})^{n_{2}+2}$ generates an iterative cascade of circles whose radii decrease geometrically, thereby completely filling the available space. Here, $R_1$ and $R_2$ are the reflections with respect to vertical lines at 0 and $a$, respectively.

In this study, we use the simplest member of the F2 family with parameters $l=4$, $n_{1}=0$, and $n_{2}=0$, corresponding to the top-left structure in Fig.~7 of Ref.~\cite{oron2000}.  According to the analysis reported in Fig.~10 of the same reference, this configuration has the fractal dimension $D= D_\mathrm{f}\approx 1.46$, placing it near the lower end of the spectrum of space-filling bearing geometries. The structure factor and mass-radius relation are shown in Fig.~\ref{figA2}. For smoothing we used \cite{cherny23} a log-normal distribution with the relative variance $\sigma_\mathrm{r}=0.25$. This procedure allows us to eliminate numerous minima and maxima that are superimposed on the decay of the structure factor.

Note that the usual Apollonian packing corresponds \cite{oron2000} to the limiting case of this generalized construction with the smallest base--loop size $l=4$ and infinite inversion depth, $(n_{1},n_{2}) \to (\infty,\infty)$.  In this
regime the inversion circles have radii $r_{1}=r_{2}=a$, and the iterative M\"obius transformations reproduce the classical Apollonian gasket~\cite{mandelbrot82} with the fractal dimension $D_\mathrm{f}=1.306\ldots$.

\begin{figure}[tb]
\centerline{\includegraphics[width=0.95\columnwidth,clip=true]{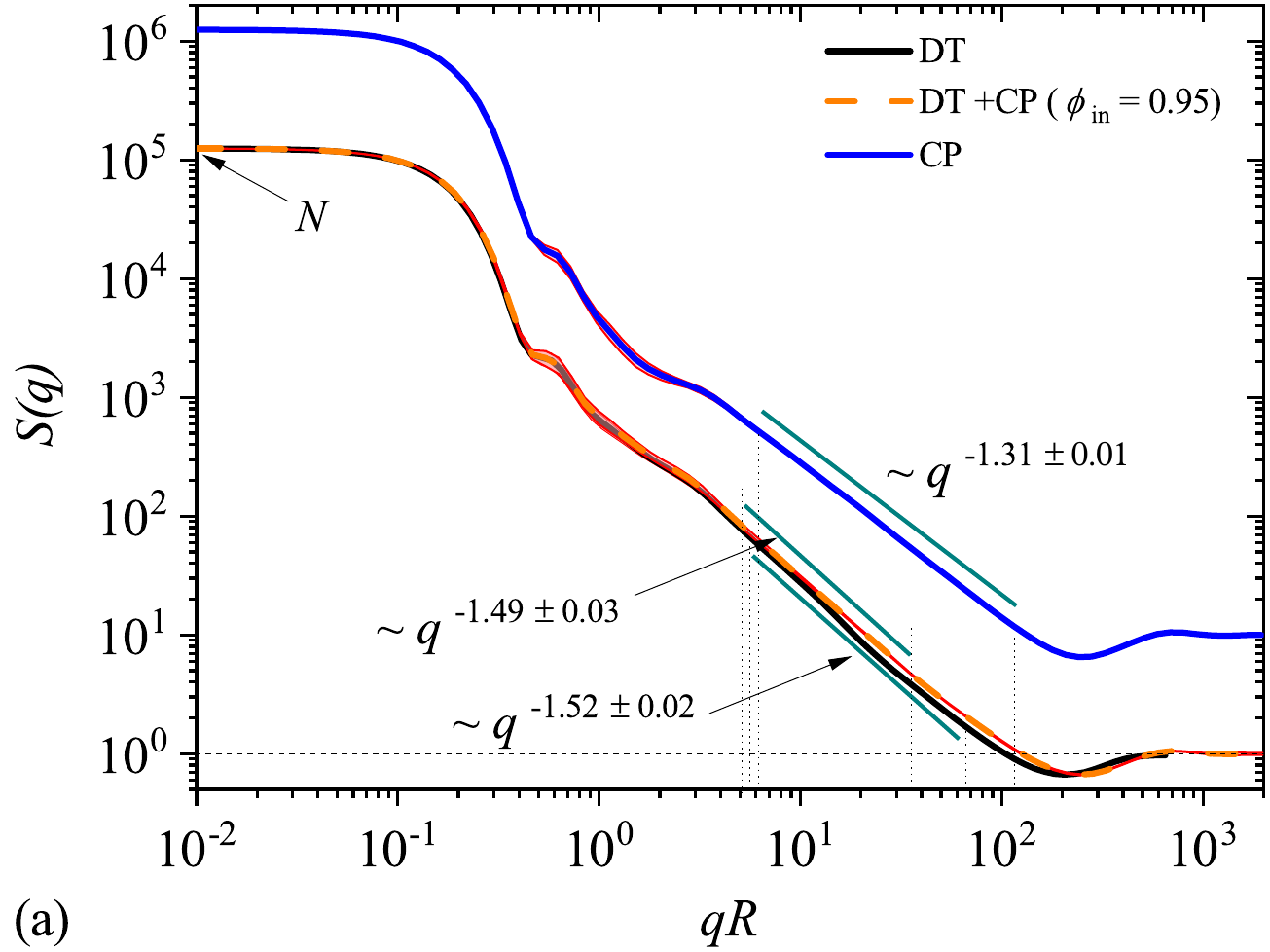}}
\centerline{\includegraphics[width=0.95\columnwidth,clip=true]{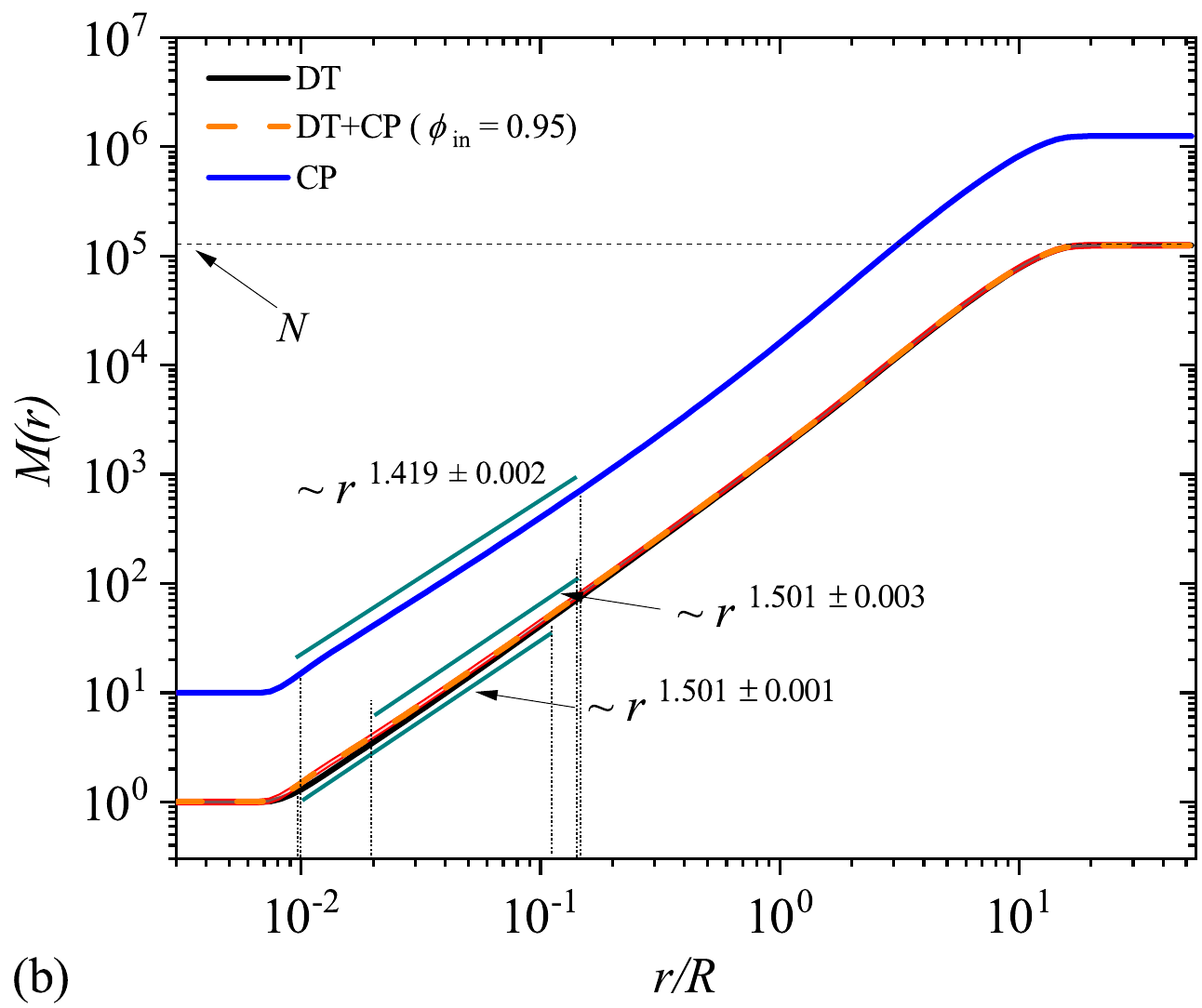}}
\caption{\label{fig:DT_CP_300} The smoothed structure factor (a) and mass-radius relation (b) for different protocols when the ratio between the largest and smallest radii is 292. The number of particles is $N = 125000$ for each protocol. The results for the DT and CP protocols are represented in solid black and blue lines, respectively. The dashed orange lines show the combination of the DT and CP protocols (see Sec.~\ref{sec:CPP300}), the corresponding curves are shifted vertically by a factor of 10 for better visualization. Vertical dotted lines denote the borders of the range over which the fit has been performed. Red curves represent errors after averaging over 20 trials.}
\end{figure}

\section{Dense random packings with DT protocol}
\label{sec:DT}

\begin{figure}[tb]
\centerline{\includegraphics[width=0.95\columnwidth,clip=true]{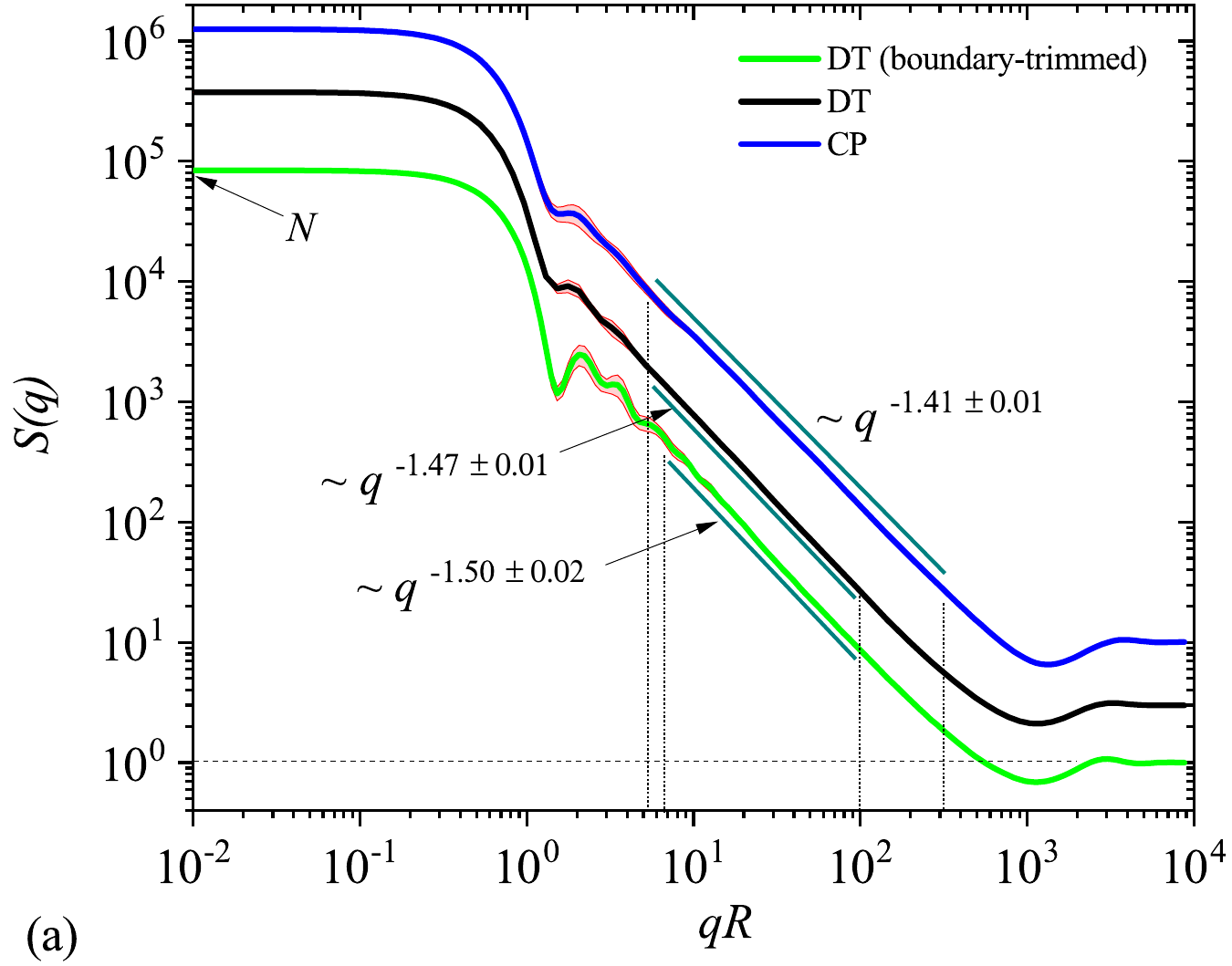}}
\centerline{\includegraphics[width=0.95\columnwidth,clip=true]{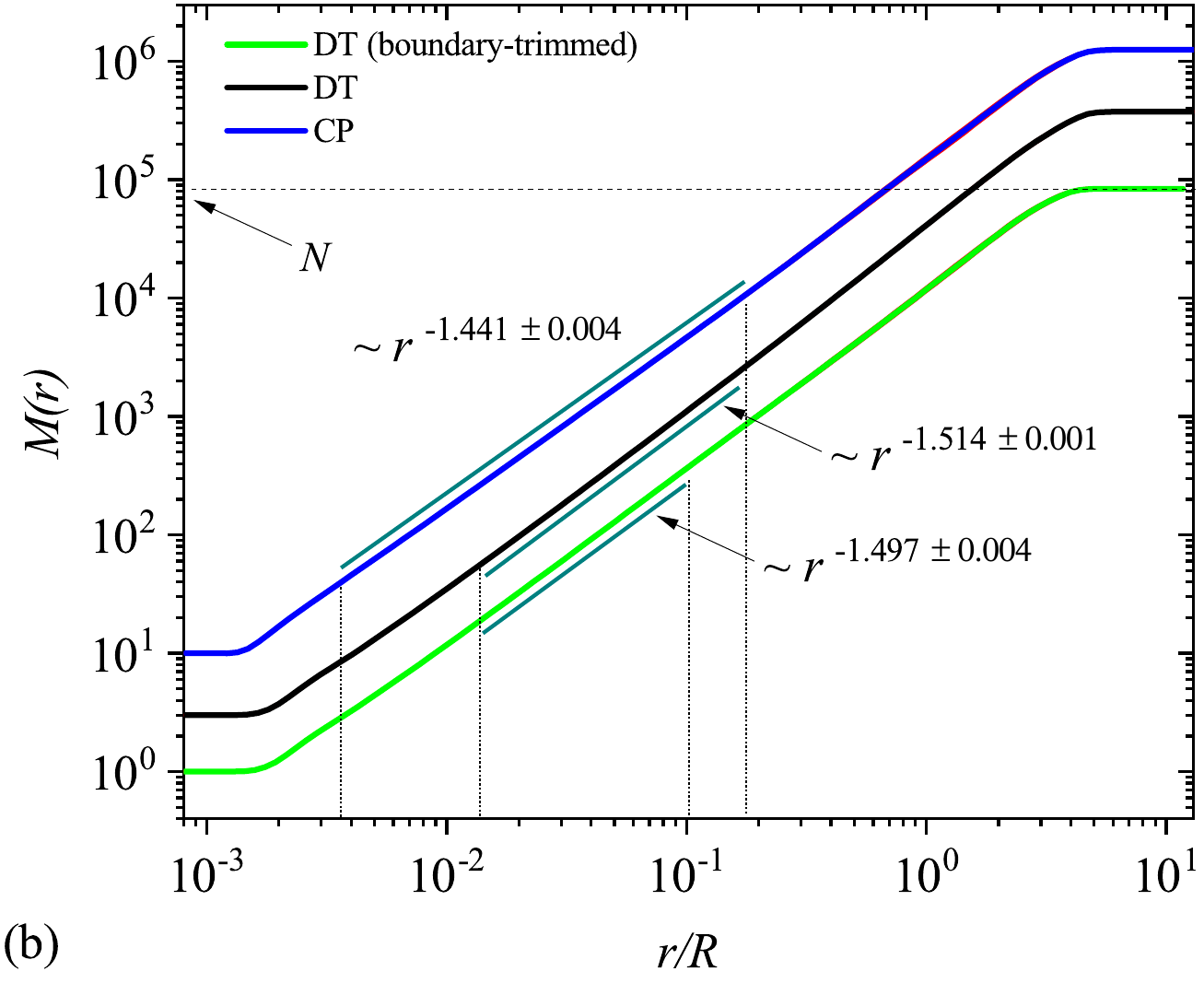}}
\caption{\label{fig:DT_CP_1500} The smoothed  structure factor (a) and mass-radius relation (b) for different protocols when the ratio between the largest and smallest radii is 1575. The results for the boundary-trimmed construction (see Fig.~\ref{figA1}) are shown in solid green lines, and for the DT and CP protocols in solid black and blue lines, respectively. The corresponding curves are  shifted vertically by a factor of 10 for better visualization. Vertical dotted lines denote the borders of the range over which the fit has been performed. Red curves represent errors after averaging over 20 trials.
}
\end{figure}

\begin{figure}[tb]
\centerline{\includegraphics[width=0.95\columnwidth,clip=true]{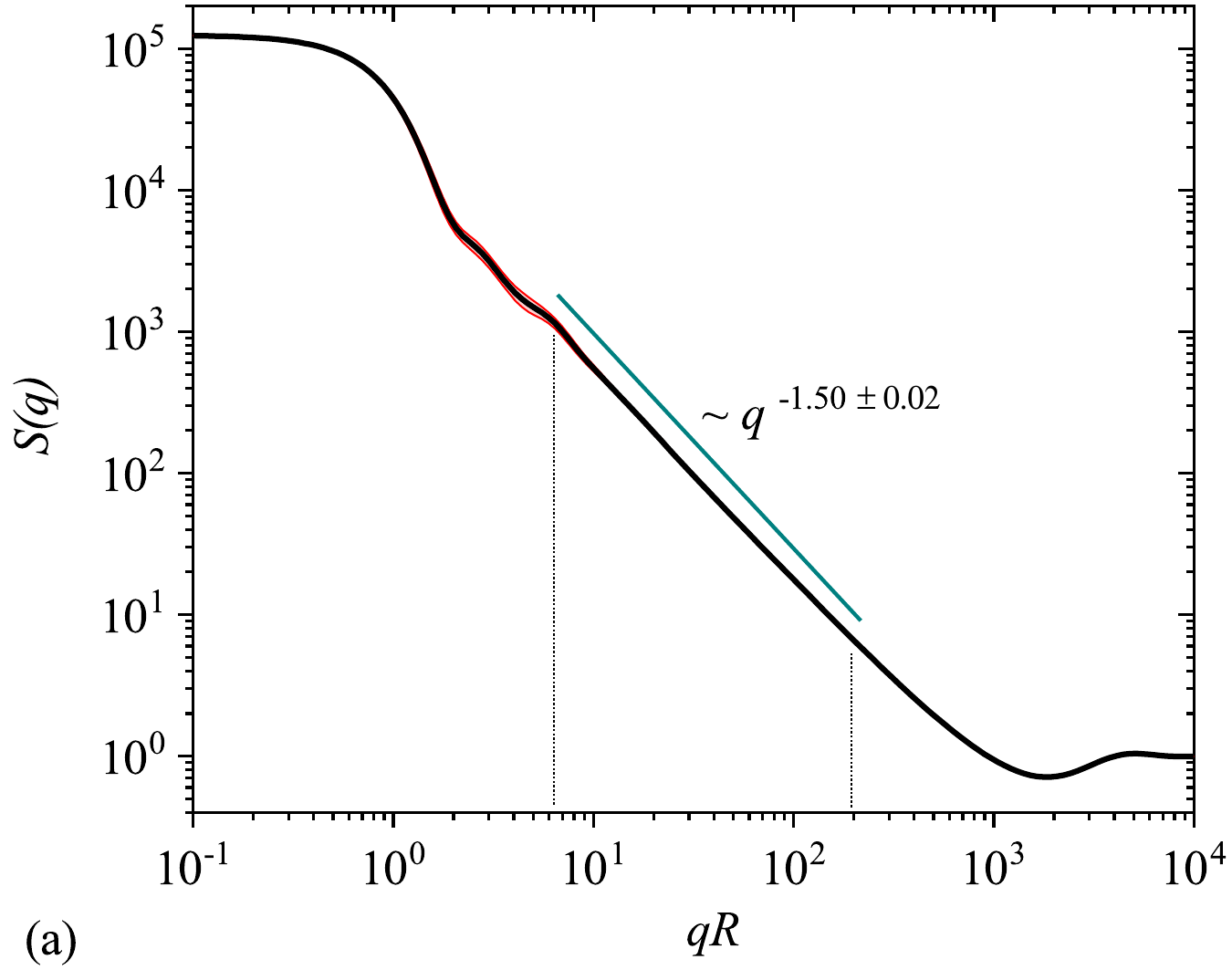}}
\centerline{\includegraphics[width=0.95\columnwidth,clip=true]{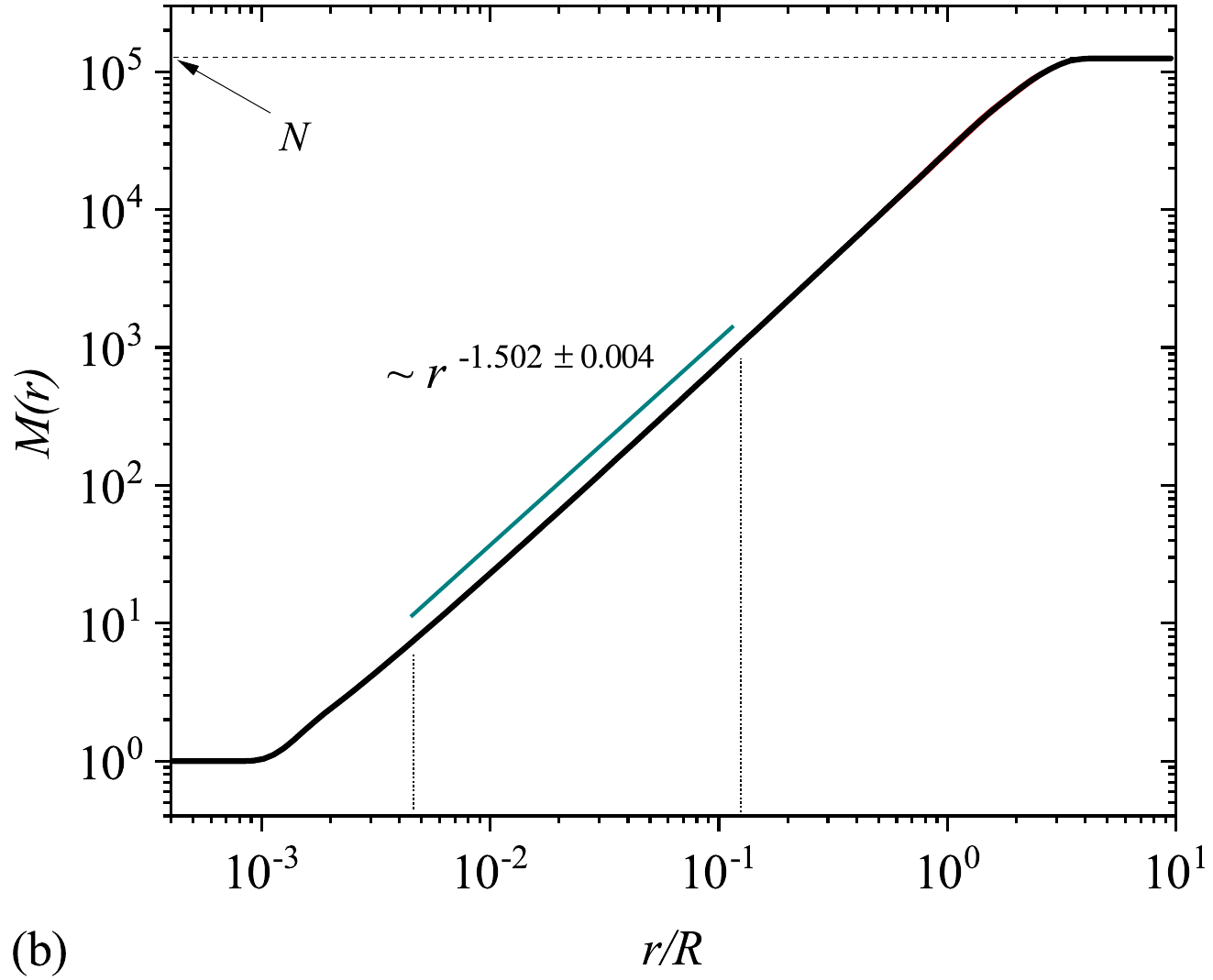}}
\caption{\label{fig:DT_CP_2500} The smoothed structure factor (a) and mass-radius relation (b) for the DT protocol with periodic boundary conditions when the ratio between the largest and smallest radii is 2500. Notations are the same as in Fig.~\ref{fig:DT_CP_300}.}
\end{figure}

To test the consistency of the fractal properties of dense random packings, they are generated using the DT protocol for different size ratios $R/a$ and boundary conditions. Figures \ref{fig:DT_CP_300}, \ref{fig:DT_CP_1500}, and \ref{fig:DT_CP_2500} show the structure factor (\ref{Sqdef}) and mass-radius relation (\ref{Mrdef}) for the total number of disks $N=125000$ and the ratios $292$, $1575$, and $2500$, respectively. Each value of $S(q)$ and $M(r)$ is obtained numerically by averaging over 20 generated trials. The smoothing function is a log-normal distribution with the relative variance $\sigma_\mathrm{r}=0.2$. The fractal ranges $2\pi/R\ll q\ll 2\pi/a$ and $a\ll r\ll R$ for $S(q)$ and $M(r)$, respectively, {widen as the size ratio increases.}

\begin{figure}[tb]
\centerline{\includegraphics[width=.7\columnwidth,clip=true]{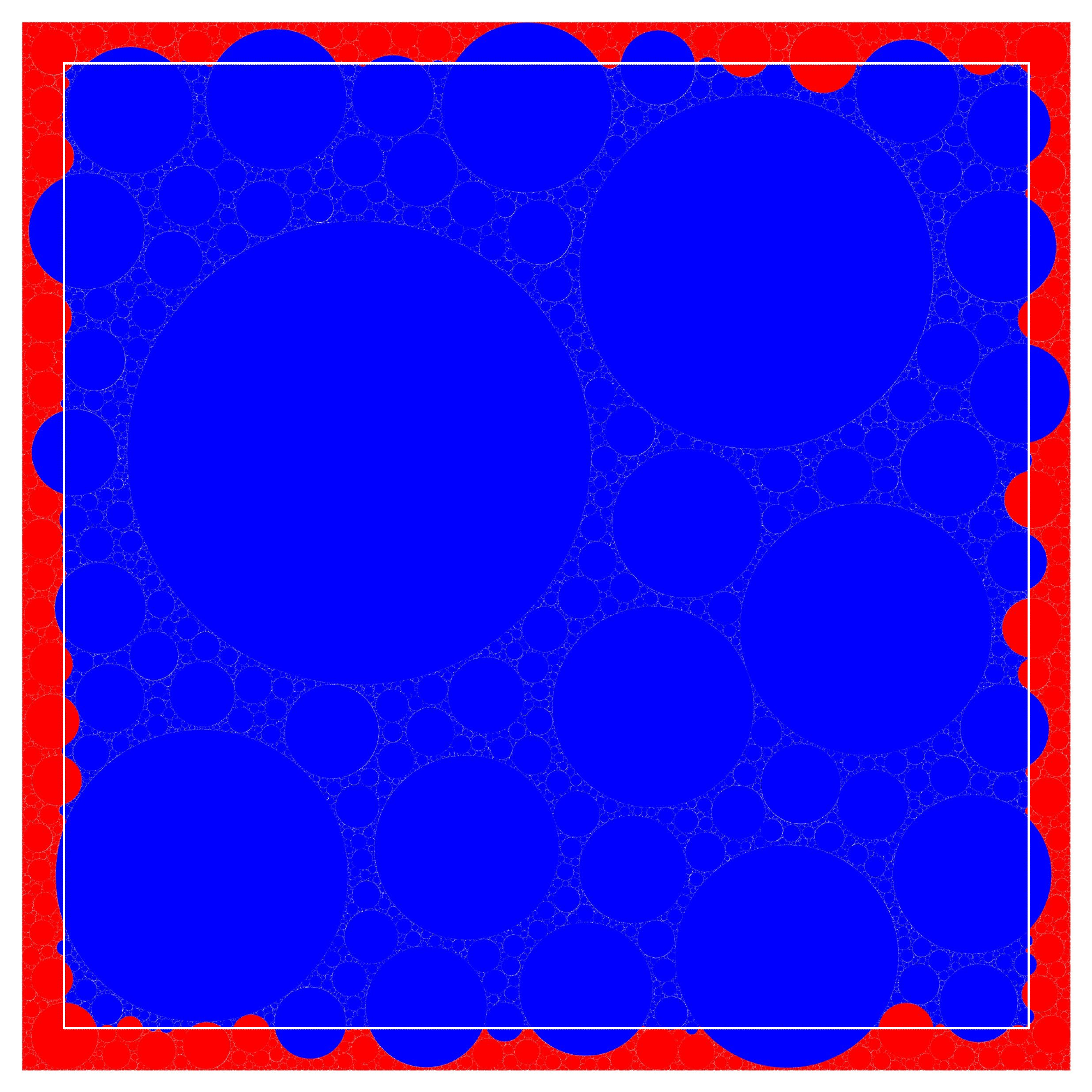}}
\caption{\label{figA1} A boundary-trimmed construction, which is obtained from the configuration generated with the DT protocol. To avoid the influence of boundary conditions, the disks (shown in red) whose centers do not fall within the internal square of edge $0.92$ have been removed. The edges of the internal square are shown as solid white lines. For each trial, the number of remaining disks is about 85000. The corresponding structure factor and mass-radius relation are represented in Fig.~\ref{fig:DT_CP_1500} (solid green lines).}
\end{figure}

Our analysis reveals (see Table \ref{tab:exponents}) that the fractal properties exhibit a high degree of consistency even for moderate values of the size ratio. The exponents of the mass-radius relation $D_{\mathrm{f}}$, the structure factor $\alpha$, and size distribution $D$ coincide very closely in accordance with the model of dense random packing developed in the previous papers \cite{cherny23} and \cite{cherny24}. {We found empirically (see Table \ref{tab:exponents}) that the coincidence of the exponents in the DT packing occurs when the packing fraction exceeds a critical value of $\sim 0.95$.} An exception is the exponents for $R/a=1575$ and zero boundary conditions (see Fig.~\ref{fig:DT_CP_1500}). They slightly deviate from each other and $D=1.5$. We explain this behaviour by the influence of zero boundary conditions: the boundaries of the square ``attract'' disks of small radii and thus make the configuration slightly inhomogeneous. To reduce the influence of the boundary effects, a part of the disks in contact with the borders of the square is removed (see Fig.~\ref{figA1}), and the structure factor and mass-radius relation are calculated for the remaining part of the system, see Fig.~ \ref{fig:DT_CP_1500}. All three exponents for the "boundary-trimmed" set of disks become very close again.

Note that the exponents, which are observed as the slopes in a double-logarithmic scale, exhibit some sensitivity to the selection of fractal ranges. Careless extension of the selected interval can change the value of the resulting exponent due to deviations from the power-law  dependence at the edges of the fractal range.

\begin{table*}
    \caption{\label{tab:exponents} The power-law exponents for the structure factors $\alpha$ [$S(q)\sim 1/q^\alpha$] and the mass-radius relation $D_{\mathrm{f}}$ [$M(r)\sim r^{D_{\mathrm{f}}}$] shown in Figs.~\ref{fig:DT_CP_300}-\ref{fig:DT_CP_2500}. The exponents are obtained from linear fits over the fractal ranges $R q_{\text{min}}\leqslant R q \leqslant R q_{\text{max}}$ and $r_{\text{min}}/R\leqslant r/R \leqslant r_{\text{max}}/R $, respectively. {$D$}, BC, and  $\phi$ denote {the exponent of the power-law radii distribution}, boundary conditions, and the packing fraction, respectively.}
    \begin{ruledtabular}
    \begin{tabular}{cccccccccccccc}
       {$D$} & Protocol& BC& $N$ & $\displaystyle{\frac{R}{a}}$ &{$\phi$} & $\alpha$& $R q_{\text{min}}$ & $R q_{\text{max}}$ & $D_{\mathrm{f}}$& $\displaystyle{\frac{r_{\text{min}}}{R}} \times 10^{-3}$ & $\displaystyle{\frac{r_{\text{max}}}{R}} \times 10^{-3}$ & Ref. \\[.5 em]
       \hline\\[-.7 em]
       {1.46} & GAP & Zero & 52366 & 1575 & 0.943 & 1.46 $\pm$ 0.02 & 15.2 & 85.1& 1.441 $\pm$ 0.002& 3.97 & 36.6 &   Fig.~\ref{figA2}\\
       {1.5}  & DT & Periodic & 125000 & 292 & 0.955 & 1.52 $\pm$ 0.02 & 5.79 & 65 & 1.501 $\pm$ 0.001 & 9.09 & 110 & Fig.~\ref{fig:DT_CP_300} \\
       {1.5}  & DT(0.95) + CP & Periodic & 125000 & 292 & 0.966 & 1.49 $\pm$ 0.03 & 5.07 & 30.41& 1.501 $\pm$ 0.003 & 19.9 & 145 & Fig.~\ref{fig:DT_CP_300}\\
       {1.5} & CP &  Periodic & 125000& 292  & 0.953 & 1.31 $\pm$ 0.01 & 6.74 & 114.65& 1.419 $\pm$ 0.002  & 10.1 & 148  & Fig.~\ref{fig:DT_CP_300}\\
       {1.5}  & DT & Zero & 125000 & 1575 & 0.983 & 1.47 $\pm$ 0.01 & 5.12 & 107.12 & 1.514 $\pm$ 0.001 & 14.0 & 177 & Fig.~\ref{fig:DT_CP_1500}\\
       {1.5}  & DT& Zero & $\sim 85000$\footnote{With $\sim 40000$ points near the square borders being removed to avoid the influence of the boundary conditions, see Fig.~\ref{figA1}.}
        & 1575 & 0.983 & 1.50 $\pm$ 0.02 & 6.94 & 99.28 & 1.497 $\pm$ 0.004 & 14.0 & 101 & Fig.~\ref{fig:DT_CP_1500}\\
        {1.5} & CP  & Periodic & 125000 & 1575 & 0.967 & 1.41 $\pm$ 0.01 & 5.12 & 310& 1.441 $\pm$ 0.004& 3.64 & 177 & Fig.~\ref{fig:DT_CP_1500}\\
        {1.5} & DT & Periodic & 125000 & 2500 & 0.987 & 1.50 $\pm$ 0.02 & 5.96 & 197& 1.502 $\pm$ 0.004& 4.22 & 121 & Fig.~\ref{fig:DT_CP_2500}\\
         {1.7} & { DT} &  { Zero} &  { 483000} &  { 2205} &  { 0.961} &  { 1.68 $\pm$ 0.02} &  { 3.71} &  { 58.55}  &  { 1.650 $\pm$ 0.003} &  { 28} &  { 333} & {SM}\footnote{{The figures can be found in the Supplementary Material.}}\\

          {1.7} & {CP}  & {Periodic} &  {124400} &  {292} &  {0.949} &  {1.43 $\pm$ 0.01} &  { 3.96} &  {46.76}   &  {1.703 $\pm$ 0.006} &  {136} &  {1976} & {SM}\\

          {1.7}  &{ CP} &  { Periodic} &  { 250000} &  { 1497} &  { 0.962} &  { 1.56 $\pm$ 0.01} &  { 4.32} &  { 105.33} &  { 1.633 $\pm$ 0.005} &  { 63} &  { 1838} & {SM}\\

          {1.9} &{ DT} &  { Zero} &  { 900000} &  { 1361} &  { 0.904} &  { 1.77 $\pm$ 0.01} &  { 2.68} &  { 56.33} &  { 1.782 $\pm$ 0.001} &  { 47} &  { 896} & {SM}\\

         {1.9}  & {CP} &  {Periodic} &  {145200} &  {292} &  {0.938}&  {1.59 $\pm$ 0.01} &  {1.13} &  {23.78} &  {1.767 $\pm$ 0.001} &  {179} &  {2073} & {SM}\\

        {1.9} & {CP} &  {Periodic} &  {250000} &  {693} &  {0.950}&  {1.66 $\pm$ 0.01} &  {2.11} &  {28.86} &  {1.760 $\pm$ 0.001} &  {144} &  {1663} & {SM}\\
         \end{tabular}
\end{ruledtabular}
\end{table*}

\section{Packing with constant pressure protocol}
\label{sec:monti}

In this section, we study the characteristics of jammed packings generated through the CP protocol for a square with imposed periodic boundary conditions. The protocol was described in Sec.~\ref{sec:disks_meth} above.

\subsection{Moderate size ratio}
\label{sec:CPP300}

{We numerically implemented the CP protocol for dense packing} with the parameters $D=1.5$ and $R/a=292$ for 20 trials.  The structure factor and mass-radius relation are shown in Fig.~\ref{fig:DT_CP_300}. The obtained fractal exponents of the mass-radius relation, $D_{\mathrm{f}}=1.419 \pm 0.002$, and the structure factor, $\alpha=1.31 \pm 0.01$, are inconsistent and significantly differ from the power-law exponent, $D=1.5$, see Table \ref{tab:exponents}. The observed discrepancies can be explained by the insufficient value of the size ratio, which is discussed in more detail in Sec.~\ref{sec:CPP1500} below. Note that the chosen size ratio $R/a$ even exceeds the $200$ value considered in the paper \cite{monti23}, for which the fractal exponent $\alpha = 1.28 \pm 0.01$ was numerically obtained at the same $D$. This value is quite close to our result.

{At a given size ratio $R/a = 292$, as the power-law exponent $D$ increases, we observe an increase in the exponent $\alpha$ (see Table \ref{tab:exponents}), in contrast to the results reported in Ref.~\cite{monti23}. We attribute this discrepancy to the smallness of the size ratio ($\sim 100$) that the authors of that paper chose. However, we found that the gap $D-\alpha$ still grows with fixed $R/a$ and increasing $D$.}

To test how jamming itself affects the features of dense packings, a combination of the DT and CP protocols is applied. At the beginning, the system is packed with the DT protocol to the higher packing fraction of $0.955$, and after that, the CP protocol is realized. The results for the combination of the packings and the single non-jammed DT packing are shown in Fig.~\ref{fig:DT_CP_300}. For both DT+CP and DT protocols, the fractal and power-law exponents agree very closely, see Table \ref{tab:exponents}. It follows from these findings that jamming itself is not a significant factor influencing the fractal behaviour of the system.

Another simple argument can be made to show the insignificance of jamming \emph{per se}. Suppose we have a perfectly jammed system of packed disks. Let us scale the system with the  factor $1+\varepsilon$ but leave the radii of the disks unchanged. It is clear that for arbitrarily small $\varepsilon > 0$ no disk is in contact with the others anymore, and then the fraction of rattlers is equal to one. On the other hand, the correlation properties, including the structure factor and mass-radius relation, remain practically unchanged.

As shown in Fig.~\ref{fig:methods}b, the packing obtained using the CP protocol contains empty cavities filled with a few rattlers. Such cavities are formed when several large particles come into contact, forming an almost empty space between them into which smaller particles can no longer penetrate.  We believe that it is the cavities and inhomogeneities that are responsible for the non-randomness of the CP packing {(see Secs.~\ref{sec:param_pores} and \ref{sec:num_pores} below)}.

\subsection{High size ratio}
\label{sec:CPP1500}

{To investigate the influence of the size ratio on the fractal exponents, CP packings were obtained with a high size ratio $R/a = 1575$ across 20 trials (see Fig.~\ref{fig:DT_CP_1500} and Table \ref{tab:exponents} for the results). The fractal exponents of the mass-radius relation, $D_{\mathrm{f}} = 1.441 \pm 0.004$, and of the structure factor, $\alpha = 1.41 \pm 0.01$, differ from those obtained for $R/a = 292$ and remain mutually inconsistent.} We conclude that the CP protocol yields inconsistent fractal behavior that depends on the initial density and particle size ratio, which leads to ambiguities in its interpretation.

Nevertheless, these results suggest that both fractal exponents $D_{\mathrm{f}}$ and $\alpha$ \emph{slowly} converge to the power-law exponent $D$ as the size ratio $R/a$ increases. The slow convergence appears to be a result of \emph{non-randomness} of the CP packings, which becomes less important at high size ratios {(see the discussion in Sec.~\ref{sec:num_pores} below)}.

However, the CP computational cost grows dramatically as the size ratio increases, which presents a practical obstacle to achieving full convergence in our modeling framework. {We note that a large size ratio increases the number of neighboring particles, which can reduce computational efficiency. The hierarchical grid algorithm \cite{ogarko12} mitigates this issue by shortening the time needed to construct a neighbor list. This algorithm is implemented in LAMMPS for parallel computations. Nevertheless,} a large ratio of the heaviest-to-lightest masses of the particles leads to an apparently high difference in their average velocities, which increases the number of time steps to reach equilibrium.

Let us estimate the number of steps in simulations. We assume that particle velocity is approximately proportional to the inverse square root of its mass, which is equal, up to a factor, to the inverse radius in two dimensions. Then we have for the highest and lowest velocities: $v_{\mathrm{max}}\sim 1/a$ and $v_{\mathrm{min}}\sim 1/R$, respectively. The time step $\tau$ should be less than $a/v_{\mathrm{max}}\sim a^2$ so that the small particles have time to interact and not fly through each other. The total simulation time should be at least $t_{\mathrm{sim}} \sim R/v_{\mathrm{min}} \sim R^2$, so that during the simulation time the largest particle has time to travel a distance not less than its radius. Then the required number of steps scales as the square of the size ratio: $t_{\mathrm{sim}}/\tau \sim R^2/a^2$.

\section{Pore size analysis}
\label{sec:pores}

\begin{table*}
    \caption{\label{tab:pores} {Parameters of the pore size distributions for different protocols (see Table \ref{tab:exponents}): the total area of the covering disks $S_\mathrm{pcov}$, one-half of the average distance between neighbour disks $d$ [Eq.~(\ref{ddef})] in units of $a$, the number of pores $N_\mathrm{p}(d)$ larger than $d$, the average radius of pores $\langle r\rangle_\mathrm{p}$ [Eq.~(\ref{rp})] in units of $d$, the area fraction of pores $S_{d}$ [Eq.~(\ref{Sa})] larger than $d$, and the exponents for the pore size distribution $D_\mathrm{p}$ and the combined pore and disk size distribution $D_\mathrm{c}$  (see the main text for details). For the CP protocol with the ratio $R/a=292$, the fractal range for the pore size distribution practically shrinks to zero\footnote{{See Fig.~\ref{fig:pores_comp} in the Supplementary Material.}}; 
    therefore, the exponent $D_\mathrm{p}$ is not given in the table. The boundary conditions are zero and periodic for the DT and CP protocols, respectively.}}
\begin{ruledtabular}
    {\begin{tabular}{cccccccccccccc} 
        $D$ & Prot. & $\displaystyle{\frac{N}{10^3}}$ & $\displaystyle{\frac{R}{a}}$ & $(1-\phi)$ &
        $S_\mathrm{pcov}$ & $\displaystyle{\frac{d}{a}}$ & $\displaystyle{\frac{N_\mathrm{p}(d)}{10^3}}$ & $\displaystyle{\frac{\langle r\rangle_\mathrm{p}}{d}}$ &
        $S_{d}$ & $D_\mathrm{p}$ & $D_\mathrm{c}$ & $\alpha$ & Ref. \\[.6 em]
       \hline\\[-.7 em]
      1.5 & DT  &  125 & 292  &  0.045 & 0.08  &0.79 & 275 & $1.71 \pm 0.009$ & $0.772 \pm 0.001$ & $1.47 \pm 0.08$  & $1.5 \pm 0.0003$ & 1.52 $\pm$ 0.02 & -- \\
      1.5 & DT  &  125 & 1575 &  0.017 & 0.03  &0.76 & 135 & $1.05 \pm 0.0008$ & $0.54 \pm 0.001$ &  $1.48 \pm 0.06$ & $1.5 \pm 0.0003$ & 1.47 $\pm$ 0.01 &Fig.~\ref{fig:methods}a \\
      1.7 & DT  &  483 & 2205 &  0.039 & 0.06  &0.96 & 246 & $0.98 \pm 0.002$ & $0.41 \pm 0.002$ &  $1.7 \pm 0.08$   & $1.7 \pm 0.0004$ & 1.68 $\pm$ 0.02 &-- \\
      1.9 & DT  &  900 & 1361 &  0.0906& 0.11  &1.06 & 420 & $0.963 \pm 0.002$ & $0.375 \pm 0.003$ &  $1.9 \pm 0.08$ & $1.9 \pm 0.0006$ & 1.77 $\pm$ 0.01 &-- \\
      1.5 & CP  &  125 & 292  &  0.047 & 0.083 &0.85 &39   & $3.9 \pm 0.3$ & $0.78 \pm 0.01$ & --                      & $1.32 \pm 0.03$ & 1.31 $\pm$ 0.01 &-- \\
      1.5 & CP  &  125 & 1575 &  0.033 & 0.06  &1.47 &22   & $16.3 \pm 1.7 $ & $0.93 \pm 0.005$ &  $1.03 \pm 0.05$      & $1.41 \pm 0.02$ & 1.41 $\pm$ 0.01 &Fig.~\ref{fig:methods}b \\
      1.7 & CP  &  124 & 292  &  0.051 & 0.08  &0.57 &75   & $2.49 \pm 0.06 $ & $0.84 \pm 0.02$ & --                    & $1.4 \pm 0.02$ & 1.43 $\pm$ 0.01 &-- \\
      1.7 & CP  &  250 & 1497 &  0.038 & 0.06  &0.65 &78   & $19.0 \pm 2.3 $ & $0.90 \pm 0.03 $&  $1.08 \pm 0.05$      & $1.52 \pm 0.03$ & 1.56 $\pm$ 0.01 &--\\
      1.9 & CP  &  145 & 292  &  0.062 & 0.08  &0.46 &95   & $1.80 \pm 0.02 $ & $0.81 \pm 0.002 $& --                 & $1.61 \pm 0.02$ & 1.59 $\pm$ 0.01 &-- \\
      1.9 & CP  &  250 & 693  &  0.05  &  0.065 &0.47 &144  & $3.0 \pm 0.5 $ & $ 0.88 \pm 0.03$ &  $1.10 \pm 0.05$      & $1.8 \pm 0.02$ & 1.66 $\pm$ 0.01 &-- \\
      1.46 & GAP &  52 & 1575 &  0.057 & 0.081 &2.1  &6.6  & 1.16 & 0.39 &  1.44                                                  & 1.47 & 1.46 $\pm$ 0.02 & Fig.~\ref{fig:methods}c \\
    \end{tabular}}
\end{ruledtabular}
\end{table*}

The main contribution to the mass-radius relation (\ref{Mrdef}) comes from the centers of the smallest disks, since their number is relatively large. The same is valid for the structure factor (\ref{Sqdef}). For instance, in the case of the power-law distribution with $R/a = 1575$ and $D=1.5$, $91\%$ of disks have a radius $(r < 5a)$. This fact leads to a paradoxical conclusion: the centers of the disks whose sizes fall within the fractal range of the mass-radius relation do not significantly contribute to this range. In fact, these disks actually create unoccupied voids for the smaller disks, and it is precisely the sizes of these voids that determine the behaviour of the mass-radius relation and structure factor in the corresponding fractal ranges. Pores always exist in a packing with a finite number of disks, and the behaviour of the structure factor is determined by the combined size distribution of the disks and pores. Figure~\ref{fig:methods}b shows the presence of relatively large cavities in the packing generated by the CP protocol. To test the hypothesis (see Sec.~\ref{sec:CPP300}) that deviations in the exponent $\alpha$ of the structure factor from $D$ (see Table~\ref{tab:exponents}) are caused by the presence of the cavities, we carefully consider the pores and propose an algorithm for approximate filling of the pores with a set of disks.

\subsection{Algorithm for approximate filling of pores by disks}
\label{sec:algorithm}

The algorithm covers pores with a set of disks, whose radii and positions are determined by the pores themselves, so it bears some analogy with the classical Apollonian packing~\cite{mandelbrot82}.

The square of unit edge length consists of two given complementary sets: mass and pores. Suppose that the pores contain some isolated sets (``islands''), in general. At the first step, we enlarge the mass by covering each of its point with a disk of small radius $\varepsilon$, that is, mathematically we build a uniform $\varepsilon$-neighbourhood of the mass. Thus the pores shrink, and some of the islands can disappear. Then we cover any disappearing island with a minimum number of disks of radius $\varepsilon$ centered on points belonging to the island. Then the procedure is repeated, and now we cover a new disappearing island in the same manner with disks of radius $2\varepsilon$ and proceed further. At the $n$th step we get a set of disks of radii $\varepsilon\leqslant r\leqslant n\varepsilon$ on the plane. The algorithm continues until the original set of pores is completely empty.

The final set of covering disks obeys some distribution $N'_\mathrm{p}(r,\varepsilon)$ depending on $\varepsilon$. Our proposition is that the distribution $N'_\mathrm{p}(r,\varepsilon)$ converges to $N'_\mathrm{p}(r)$ as $\varepsilon \to 0$ for any $r \geqslant r_0$, where $r_0$ is any fixed lower-cutoff radius. In practice, the limit $\varepsilon\to 0$ is not needed, it is sufficient to choose a finite $\varepsilon\ll r_0$. Thus we are left with a finite set of covering disks with the size distribution $N'_\mathrm{p}(r)$ and radii ranging from {$r_0$} to a certain maximum radius $r_\mathrm{pmax}$. In general, the covering disks may overlap each other, and, hence, their total area $S_\mathrm{pcov}=\pi \sum_{i}r_{i}^{2}$ may exceed the area of the original pores, which is equal to $1-\phi$. So, we need to normalize $N'_\mathrm{p}(r)$ to make the areas equal. Since $-{d N'_\mathrm{p}(r)}/{d r} =\sum_{i}\delta(r-r_{i})$, the area of the covering disks can be calculated as
\begin{align}\label{area_cov}
S_\mathrm{pcov}= -\pi\int_{0}^{\infty} dr \frac{d N'_\mathrm{p}(r)}{d r}r^2=2\pi\int_{0}^{r_\mathrm{pmax}} dr N'_\mathrm{p}(r) r.
\end{align}
Then the normalization factor is given by $F_\mathrm{p} = S_\mathrm{pcov}/(1-\phi)$, and we arrive at the final distribution of the covering disks $N_\mathrm{p}(r) = N'_\mathrm{p}(r)/F_\mathrm{p}$. We refer to this distribution of covering disks as the pore size distribution.

Note that this algorithm, applied to the mass, accurately reproduces the polydisperse distribution of the disks $N(r)$, since for each isolated disk it gives a disk of the same radius and in the same position. For an arbitrary isolated convex shape, the algorithm outputs the center and radius of the inscribed circle. This also implies that this disk does not cover the convex shape completely. On the other hand, the set of pores is essentially non-convex, and these cases are statistically insignificant.

\subsection{{Characteristic parameters of pores and randomness of packing}}

\label{sec:param_pores}

{The algorithm described above enables us to effectively represent pores as a set of disks and obtain the distribution of their sizes. Here we focus on characteristic parameters of the distribution. First, the distribution $N_\mathrm{p}(r)$ may contain a fractal range with a power-law exponent $D_\mathrm{p}$, but this fractal region practically shrinks to zero when the size ratio is small (see Sec.~\ref{sec:num_pores} below).}

{Second, for any size ratio, one can introduce the non-normalized distribution of areas of the covering disks:
\begin{align}
\rho_{\mathrm{p}}(r)=\pi r^2 \left(-\frac{d N_\mathrm{p}(r)}{d r}\right)=\sum_{i}\pi r_i^2 \delta(r-r_{i})
 \label{Spore_dist}
\end{align}
This distribution compensates for the substantial difference between the number of small and large pores because, although the number of small pores dominates, their contribution to $\rho_{\mathrm{p}}(r)$ is reduced by the prefactor $\pi r^2$.}

{One can calculate the average radius of pores $\langle r\rangle_{\mathrm{p}}$ weighted by pore area, where $\langle \cdots\rangle_{\mathrm{p}}$ stands for the average with respect to the normalized distribution $\rho_{\mathrm{p}}(r)/\int_0^{\infty} \d r \rho_{\mathrm{p}}(r)$. We thus obtain
\begin{align}\label{rp}
\langle r\rangle_\mathrm{p}=\frac{\int_0^{\infty} \d r \rho_{\mathrm{p}}(r)\,r}{\int_0^{\infty} \d r \rho_{\mathrm{p}}(r)}=\frac{\sum_i r_i^3}{\sum_i r_i^2},
\end{align}
where $i$ is the pore index.}

{Another parameter is the area fraction of large pores whose radii exceed one-half of the average minimum distance between disks, $d$:
\begin{equation}\label{Sa}
S_{d} = \frac{\int_{d}^{\infty} \d r \rho_{\mathrm{p}}(r)}{\int_0^{\infty} \d r \rho_{\mathrm{p}}(r)}=\sum_{\substack{i\\r_i>d}} r_i^2 \left/ \sum_i r_i^2\right.,
\end{equation}
where, in the first sum, the index $i$ runs over pores with radii larger than $d$. The parameter $d$ is one-half of the average distance between neighbouring disks, which is estimated as
\begin{align}\label{ddef}
 d=2\frac{1-\phi}{L_\mathrm{tot}}
\end{align}
with $1-\phi$ and $L_\mathrm{tot}=2\pi\sum_{i=1}^{N}r_{i}$ being the total area of pores and total circumference of the packed disks, respectively. The prefactor depends on the shape of the area, and the chosen value $2$ corresponds to a disk or square. For this reason, Eq.~(\ref{ddef}) provides an estimation rather than an exactly defined quantity.
}

\begin{figure}[bt]
\centerline{ \includegraphics[width=\columnwidth,clip=true]{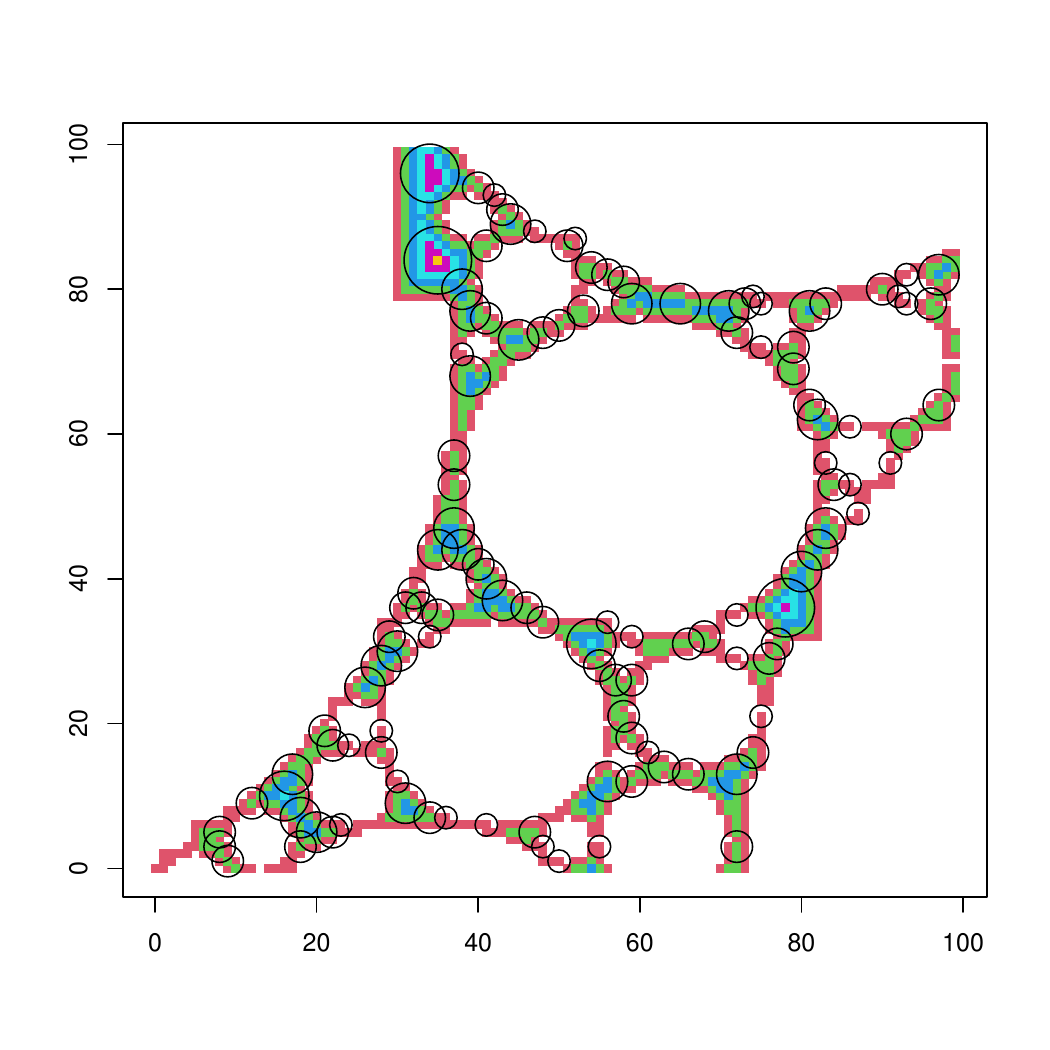}}
\caption{\label{fig:algor} Pore size distribution: graphical illustration of the algorithm (Sec.~\ref{sec:algorithm}), implemented in pixels (Sec.~\ref{sec:num_pores}). The mass increments are shown in red, green, blue, cyan, magenta, and orange, which correspond to the step numbers $1$, $2$, $3$, $4$, $5$, and $6$, respectively.
}
\end{figure}

{The above parameters (\ref{rp}) and (\ref{Sa}) are related to the randomness of packing: the smaller $\langle r\rangle_\mathrm{p}/d$ and $S_{d}$, the higher the randomness of packing. The relation to randomness can be explained by the following example. Suppose we have an infinite number of randomly packed disks that completely fill a square (perfect packing). When all disks of radii $r_i < a$ are removed, the radii of the resulting pores are indeed smaller than $a$ with very high probability, because in random filling almost every disk touches at least three disks of larger radii. Consequently, the appearance of pores with radii larger than $a$ is statistically insignificant and, if realized, indicates non-randomness of the packing\footnote{{We emphasize that randomness of a dense packing is a sufficient but not necessary condition for the absence of large pores. For instance, in the deterministic classical Apollonian packing, there are no pores with radii $r_i > a$ at each step of the algorithm.}}. However, this reasoning applies to perfect filling. When the packing fraction does not reach its maximum possible value, we need to use another parameter, $d$ (\ref{ddef}), which involves an appropriate renormalization. An increase in the parameters $\langle r\rangle_\mathrm{p}/d$ and $S_{d}$ indicates non-randomness, that is, a bias in the spatial distribution of packed disks.
}

\subsection{Numerical analysis of the pores}
\label{sec:num_pores}

\begin{figure}[bt]
\centerline{\includegraphics[width=\columnwidth,clip=true]{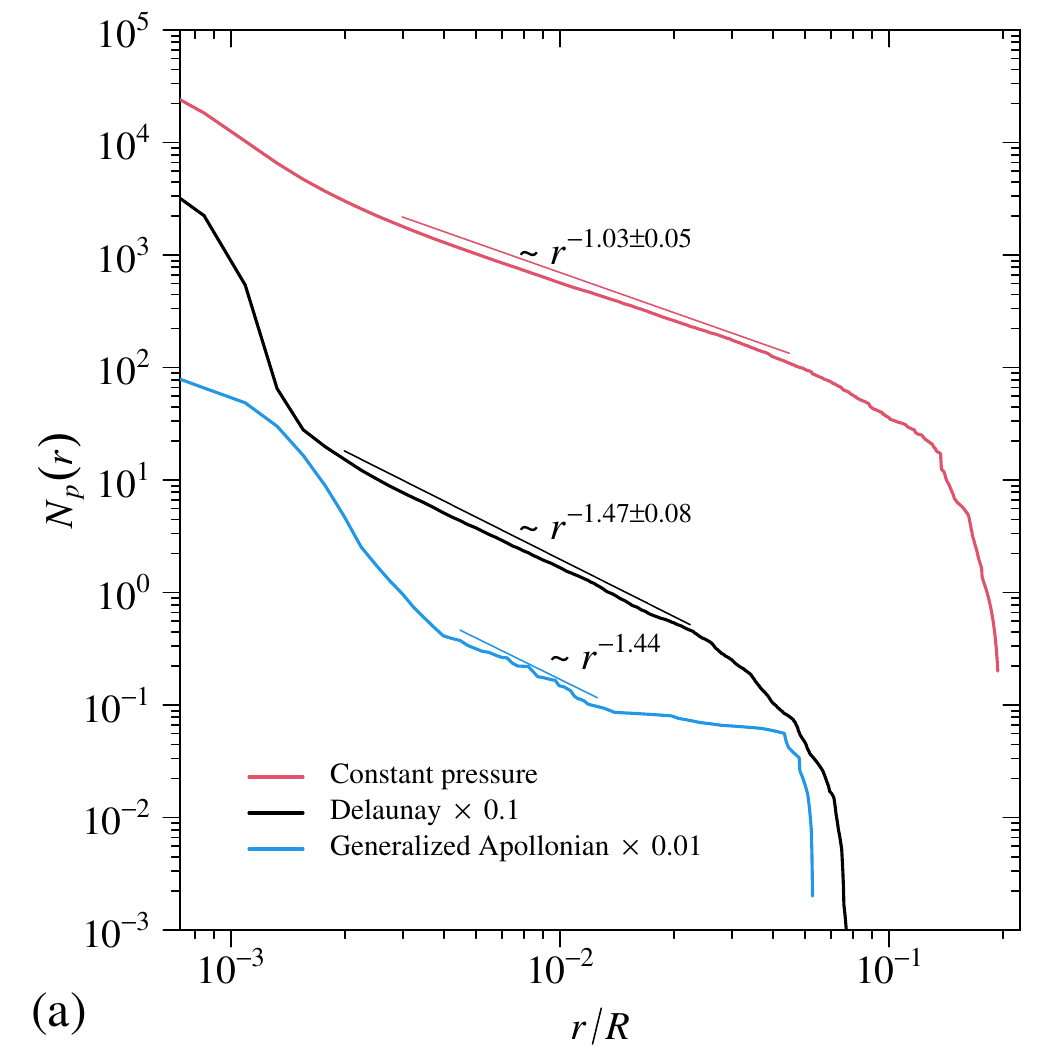}}
\centerline{\includegraphics[width=\columnwidth,clip=true]{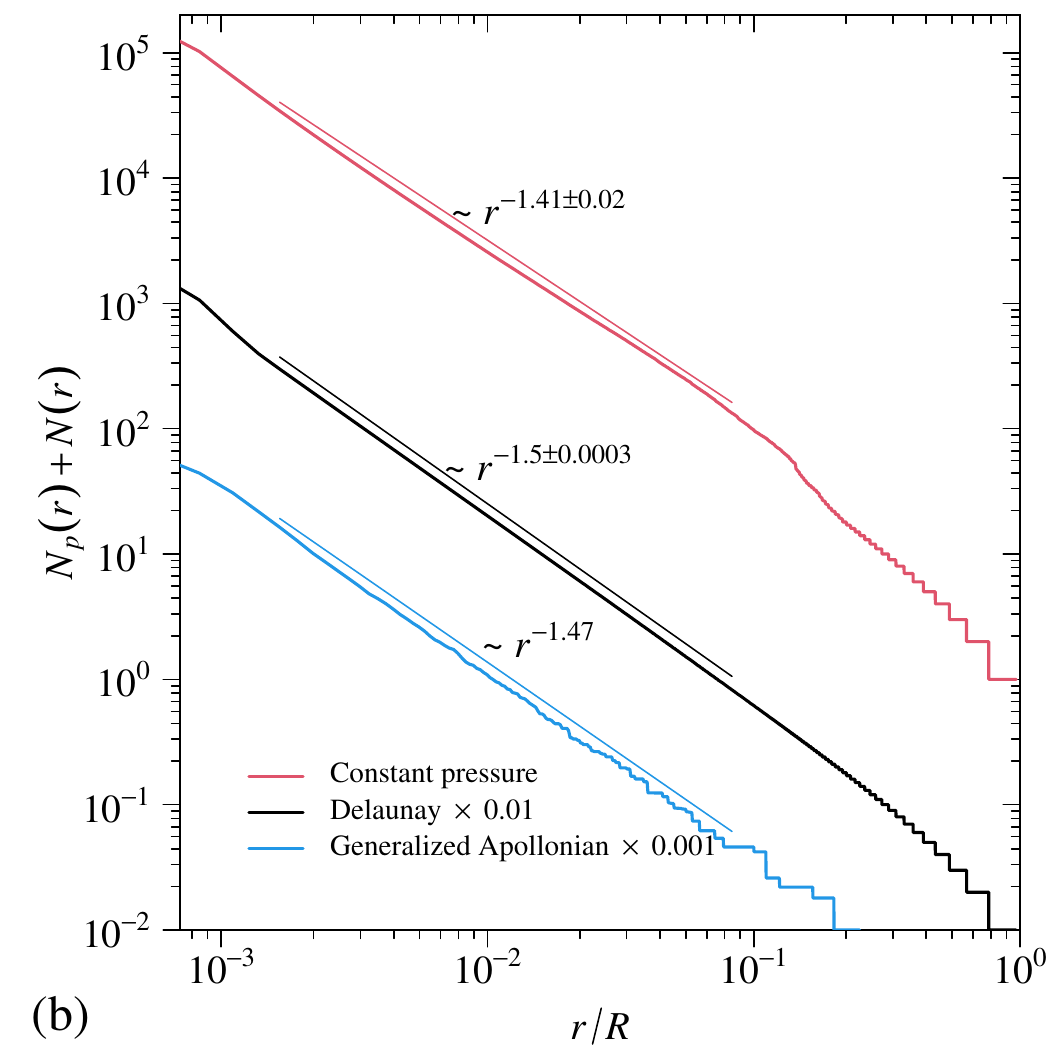}}
\caption{\label{fig:pores} The cumulative size distributions of pores {for $r\geqslant a$} (see Sec.~\ref{sec:pores} for details): $N_\mathrm{p}(r)$ (a) and the combined pores and disks $N(r) + N_\mathrm{p}(r)$ (b). The distributions for the CP protocol are shown in red, for DT in black, and for generalized Apollonian in blue. The curves are shifted vertically by a factor of 10 for better visualization.}
\end{figure}

For practical computations, a discrete version of the above algorithm is used. We represent the pores on a pixel grid $16384 \times 16384$ with the pixel size $\varepsilon =1/16384 \simeq 6.1\times 10^{-4}$.  {The ratio $a/\varepsilon$ varies from 5 to 7 across different packings.} We assign 1 to pixels belonging to the pores (filled pixels) and 0 to the rest (empty pixels). At each step we assign 0 to pixels bordering zero pixels. If a removed pixel is isolated, i.e., all its neighbors are zero or have become zero at the current step, we record that pixel as the center of a pore with radius $r = n \varepsilon$, where $n$ is the step number. After normalization, this yields a discrete pore size distribution $N_\mathrm{p}(r)$. Several steps of this algorithm are shown schematically in Fig.~\ref{fig:algor}.

{When the size ratio is sufficiently high, the distribution $N_\mathrm{p}(r)$ exhibits a power-law behaviour with an exponent $D_\mathrm{p}$ in some fractal range (see Fig.~\ref{fig:pores}a).} In the generalized Apollonian and Delaunay cases, the exponents $D$ of the distribution of disk sizes and $D_\mathrm{p}$ coincide, as expected for a uniform random filling. In the CP case, the exponents are different: $D_\mathrm{p} \simeq 1$ while $D = 1.5$.

Figure \ref{fig:pores}b represents the combined distribution $N(r) + N_\mathrm{p}(r)$, consisting of both disks and pores, in which the number of pores is normalized by the total area of the pores (see the discussion in Sec.~\ref{sec:algorithm} above). Since $D_\mathrm{p}\simeq D = 1.5$ in the generalized Apollonian and Delaunay packings, the exponent of the combined distribution is equal to $D$ as well. By contrast, in the CP packing, the effective exponent of the combined distribution becomes $D_\mathrm{c} =1.41 \pm 0.02$ in the fractal range. This coincides with the exponent $\alpha$ of the structure factor (see Table~\ref{tab:pores}) and confirms the role of cavities in the {pore structure in the reduction} of this exponent. Note that the normalization factor $F_\mathrm{p}$ for the pore size distribution (see Sec.~\ref{sec:algorithm} above) is important for the power-law exponent of the combined distribution $N(r) + N_\mathrm{p}(r)$.

{The parameters of the resulting pore size distributions across different packings are summarized in Table~\ref{tab:pores}. At a given distribution of radii of packed disks, the smaller the parameters $S_{d}$, and $\langle r\rangle_\mathrm{p}/d$ [see Eqs.~(\ref{Sa}) and (\ref{ddef})], the smaller the sizes of pores and the more random the packing (see the discussion at the end of Sec.~\ref{sec:param_pores} above). First, we observe that the CP protocol generates pores of larger sizes, compared to the DT protocol with the same number of particles and size ratio. Since the number of very large pores in the CP packing is small anyway, its spatial distribution looks inhomogeneous due to fluctuations (see  Fig.~\ref{fig:methods}). Thus, the DT protocol is more homogeneous and more random, compared to the CP protocol.}

{Second, for the DT protocol, at a given exponent $D$ of the size distribution of packed disks, the larger the size ratio $R/a$, the smaller the relative average pore radius $\langle r\rangle_\mathrm{p}/d$ and the area fraction of large pores $S_{d}$, and hence the more random and better the packings\footnote{{In the DT packing with $D=1.9$, the exponents $\alpha$ and $D_\mathrm{f}$ are smaller than $D$ due to insufficient density of the packing.}}. For the CP protocol, we observe the opposite tendency. Nevertheless, for the CP protocol, the exponent $\alpha$ becomes closer to $D$ and the packing fraction increases as the size ratio $R/a$ grows. This apparent paradox can be resolved by a simultaneous decrease in the ratio $N_\mathrm{p}(d)/N$: although the sizes of pores grows, their relative number diminishes and, hence, their contribution to the decrease in the exponent $\alpha$ within the fractal range of the structure factor becomes smaller.
}

{Third, at a given size ratio and a given number of packed disks, increasing $D$ improves the randomness of packing for the CP protocol but reduces it for the DT protocol. Nevertheless, for the CP packing, the gap $D-\alpha$ grows as $D$ increases (see Table \ref{tab:pores} for the ratio $R/a=292$ and various $D$). While the area fraction $S_{d}$ remains approximately the same ($S_{d}\approx 0.8$) and $\langle r\rangle_\mathrm{p}/d$ decreases ($\langle r\rangle_\mathrm{p}/d=3.9, 2.5, 1.8$ for $D=1.5, 1.7, 1.9$, respectively), the ratio $N_\mathrm{p}(d)/N$ grows. We conclude that it is the relative number of large pores $N_\mathrm{p}(d)/N$ that controls the deviation of the exponent $\alpha$ from the exponent $D$ of the size distribution of packed disks.
}

\section{Conclusions}
\label{sec:concl}

The model of dense packing of a specified set of disks states \cite{cherny23,cherny24} that all three exponents $D_{\mathrm{f}}$, $\alpha$, and $D$ (for the mass-radius relation, structure factor, and power-law distribution, respectively) coincide if the packing is random, the density is high, and the size ratio $R/a$ is sufficiently large. Deviations from this behaviour indicate a violation of at least one of the above conditions.

The results obtained agree with this statement, see Table \ref{tab:exponents}. The DT and RSA protocols are random by construction, and if the other conditions are fulfilled then indeed we have $D_{\mathrm{f}}=\alpha=D$ within computational error. {The key parameter, responsible for the coincidence of the exponents, is the packing fraction, whose critical value is found to be about $\phi_{\mathrm{c}}\simeq 0.95$. When $\phi> \phi_{\mathrm{c}}$, the exponents coincide}.

The exponents for the CP protocol are inconsistent and lack universality, since the exponents depend on the initial packing conditions and the size ratio (see Sec.~\ref{sec:CPP300}). The inconsistency can be explained by the presence of relatively large cavities in the packing (see Fig.~\ref{fig:methods}b), while jamming per se has no significant effect on the fractal properties (see the discussion in Sec.~\ref{sec:CPP300}). The effect of the cavities is twofold: it reduces both packing fraction and randomness of the CP packings.

To study the effect of cavities, we developed an algorithm to obtain the pore size distribution (see Sec.~\ref{sec:pores} and Table~\ref{tab:pores}). The combined distribution $N(r) + N_\mathrm{p}(r)$ of both disks and pores obeys a power law with exponent $D_\mathrm{c}$. Within numerical error, it coincides with the fractal exponent $\alpha$ of the structure factor for {almost all the packings} studied.

The numerical results reveal a clear tendency of the fractal exponents to approach $D_\mathrm{f} = \alpha = D$ with increasing size ratio, pointing to a common asymptotic behaviour for different packing protocols. {For the CP protocol, the exponent $\alpha$ becomes closer to $D$ as the size ratio increases (see Table \ref{tab:exponents}); however, the gap $D-\alpha$ grows with fixed $R/a$ and increasing $D$. It is found (see Sec.~\ref{sec:num_pores}) that the relative number of large pores $N_\mathrm{p}(d)/N$ controls the deviation of the exponent $\alpha$ from the exponent $D$ in the CP protocol: the larger $N_\mathrm{p}(d)/N$, the greater the deviation $D - \alpha$. We also conclude that the closer $D$ is to $2$, the higher $R/a$ must be to make $\alpha$ closer to $D$.}

The presence of the large cavities is directly related to the randomness of the CP packing, since it implies its lower configurational entropy and, consequently, its nonrandomness. This can be easily understood in the pixel representation, where large blocks accumulate a large number of filled pixels, and thus the number of ways to arrange the remaining filled pixels decreases. {A direct calculation of the configurational entropy is rather difficult.}

Another way of quantifying randomness is to introduce an appropriate order parameter. More recently, a disorder criterion was formulated for a similar task -- dense random packing of disks of identical radii \cite{Blumenfeld21}. {We introduced several parameters to describe the randomness of a packing: one-half of the average distance between neighboring disks, $d$ [Eq.~(\ref{ddef})]; the average radius of pores, $\langle r\rangle_\mathrm{p}$ [Eq.~(\ref{rp})]; the area fraction of pores larger than $d$, $S_{d}$ [Eq.~(\ref{Sa})], and the relative number of large pores $N_\mathrm{p}(d)/N$. We argue (see Sec.~\ref{sec:param_pores} above) that the smallness of $\langle r\rangle_\mathrm{p}/d$ and $S_{d}$ indicates the randomness of packings. However, the parameter $d$ is defined only up to a prefactor, which does not allow us to define the order parameter precisely. This task is a challenging problem that opens up new perspectives for research.}

\appendix

\section{General definitions and relations}
\label{sec:gen_rel}

In this appendix we follow Section II of our previous paper \cite{cherny23}, where the definitions and relations are explained in more detail.

For a set of $N$ points of the unit weight located at the positions $\bm{r}_{1},\cdots,\bm{r}_{N}$, the mass-radius relation $M(r)$ is defined as the average value of mass enclosed in the imaginary circle of radius $r$, which is centered on a point belonging to the set \cite{gouyet96:book}. According to the definition, it is given by
\begin{align}
M(r)=\frac{1}{N}\sum_{i,j}\theta(r-r_{ij})=1+\frac{1}{N}\sum_{i\neq j}\theta(r-r_{ij}), \label{Mrdef}
\end{align}
where $r_{ij}=|\bm{r}_{i}-\bm{r}_{j}|$ and  $\theta(z)$ is the Heaviside step function, that is, $\theta(x)=1$ for $x\geqslant0$ and zero elsewhere. Then $M(r)=1$ when $r$ is less than the smallest distance between points and $M(r)=N$ when $r$ exceeds the largest distance. A similar definition of the mass-radius relation was used in Ref.~\cite{Grassberger83}.

The structure factor of the set of points is defined as \cite{Cherny2011}
\begin{align}\label{Sqdef}
S(q)=\frac{1}{N}\left\langle\rho_{\bm{q}}\rho_{-\bm{q}}\right\rangle_{\hat{q}},
\end{align}
where $\rho_{\bm{q}}=\sum_{j}e^{-i\bm{q}\cdot\bm{r}_{j}}$ is the Fourier transform of the density of the points $\rho(\bm{r}) =\sum_{j} \delta(\bm{r}-\bm{r}_{j})$, and the brackets $\langle\cdots\rangle_{\hat{q}}$ stand for the average over all directions of unit vector $\hat{q}$ along $\bm{q}$. By definition, the structure factor depends only on the absolute value of $\bm{q}$, which we denote as $q$. It obeys the conditions $S(q)\simeq 1$ when $q\to\infty$ and $S(q)=N$ at $q=0$. The structure factor can be measured in small-angle scattering experiments \cite{Teixeira1988Small-angleSystems}.

In two dimensions, the mass-radius relation and the structure factor are related by the following equation
\begin{align}
\frac{1}{r}\frac{\partial M}{\partial r}=\,&\frac{1}{2\pi}\int d^2\!q\,e^{i\bm{q}\cdot\bm{r}}[S(q)-1]\nonumber\\
=&\int_{0}^{\infty} dq\, q J_{0}(q r) [S(q)-1],
 \label{relg_S_M}
\end{align}
where $J_{0}(z)$ is the Bessel function of zeroth order. When $S(q)-1\sim 1/q^{D_{\mathrm{f}}}$ for $q\to\infty$ and the exponent lies within the range $1<D_{\mathrm{f}}<2$, then its two-dimensional Fourier transform $\frac{1}{r}\frac{\partial M}{\partial r}$ is proportional \cite{erdelyi56:book} to $1/r^{2-D_{\mathrm{f}}}$ for $r\to0$. It follows that $M(r)\sim r^{D_{\mathrm{f}}}$ at sufficiently small values of $r$. Note that for the densely packed set of disks described in Sec.~\ref{sec:disks_meth}, such infinite-range asymptotics of the structure factor can be realized only in the limit of the infinite size ratio: $R/a\to\infty$ and {$R=\text{const}$}.

\bibliography{references}

\newpage
\onecolumngrid

{\section*{Supplementary Material}}

This supplementary material contains supporting numerical results for the $D = 1.7$ and $D = 1.9$ packings, including representative configurations, structure factors $S(q)$, and mass-radius relations $M(r)$. The supplementary material also provides a comparison of pore-size and combined pore-and-disk size distributions for CP packings with $D = 1.5$ at different size ratios.

\begin{figure}[htbp]
\centering
\begin{tabular}{ccc}
\includegraphics[width=0.28\textwidth,clip=true]{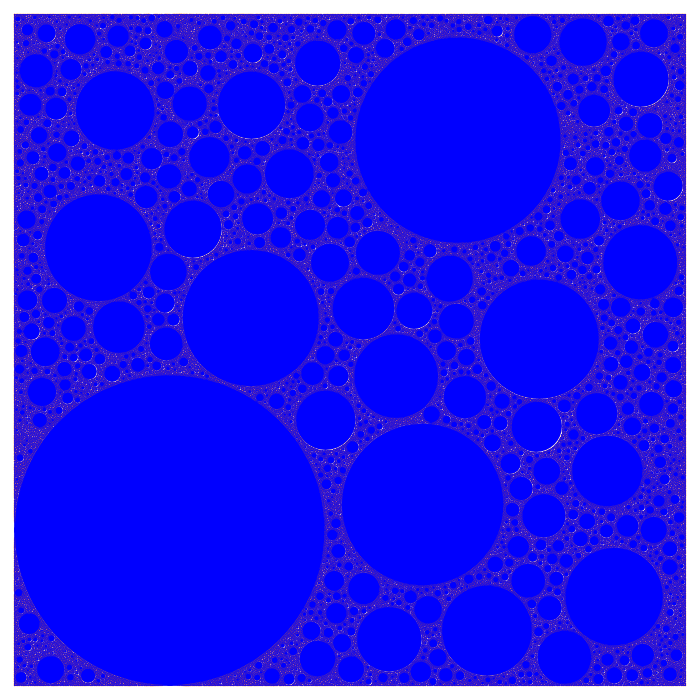} &
\includegraphics[width=0.33\textwidth,clip=true]{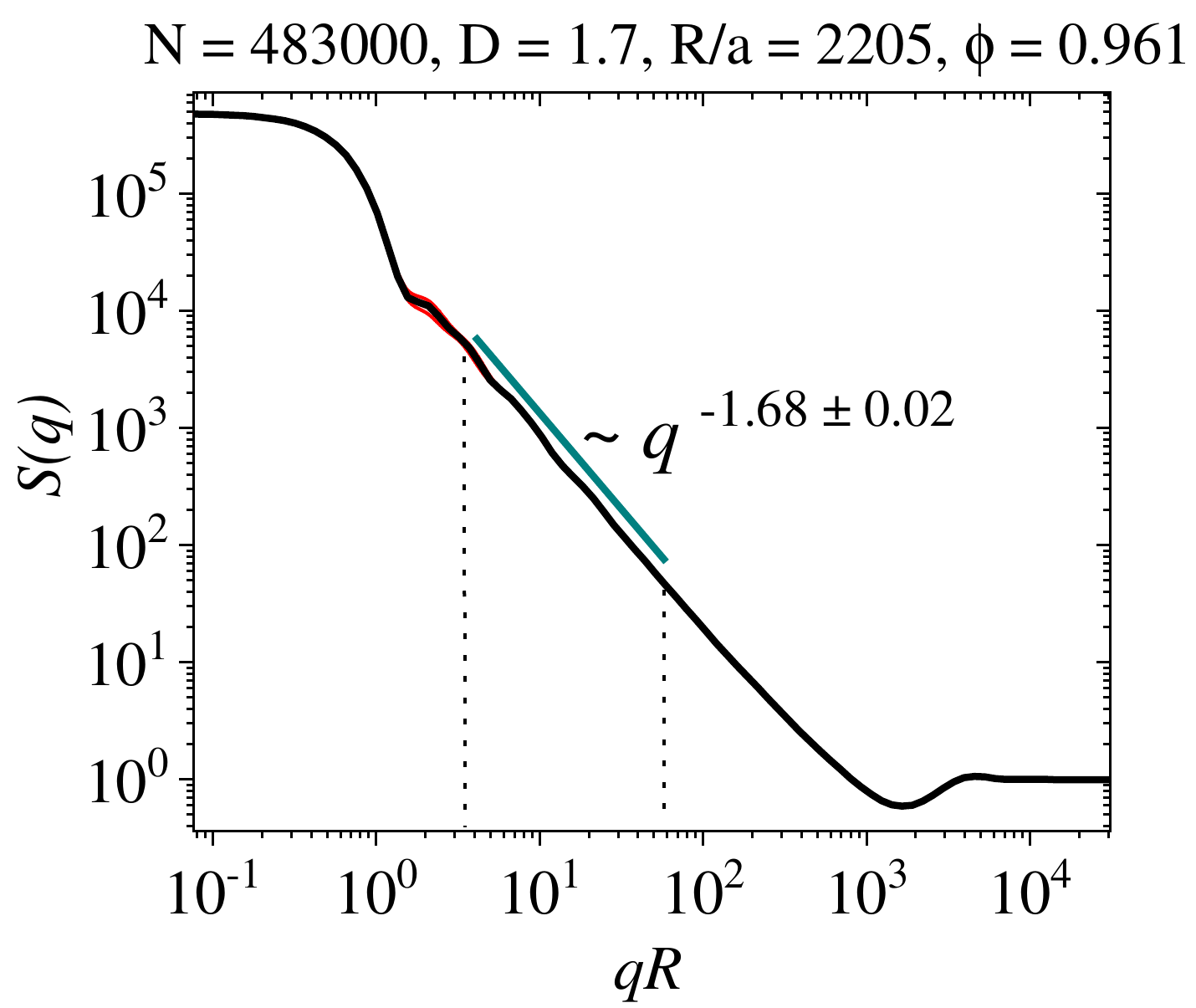} &
\includegraphics[width=0.33\textwidth,clip=true]{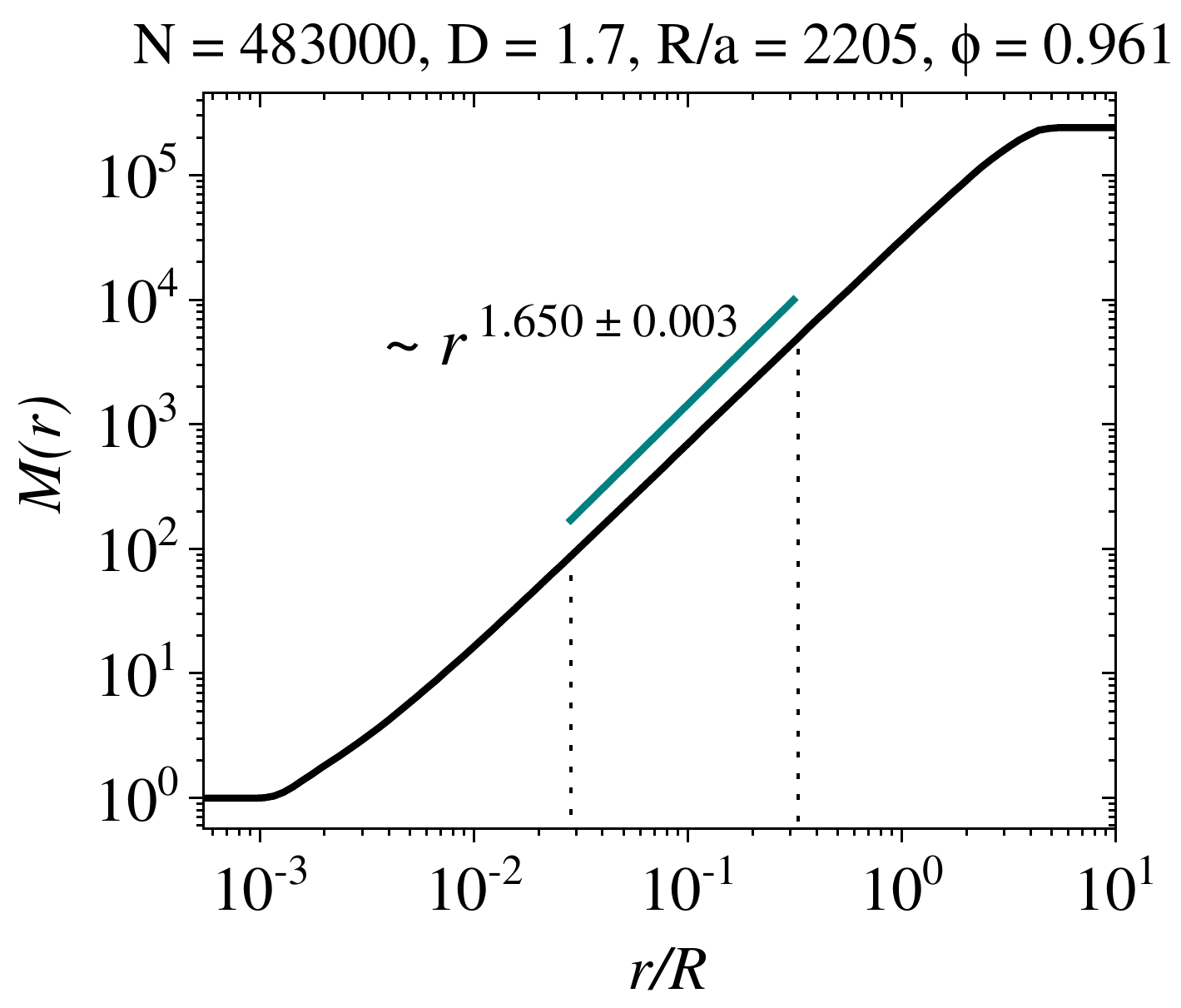} \\[4mm]

\includegraphics[width=0.28\textwidth,clip=true]{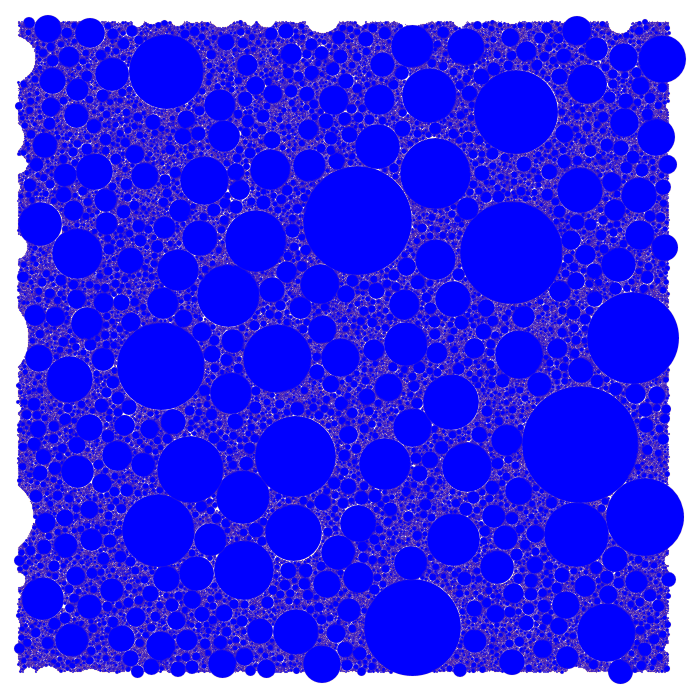} &
\includegraphics[width=0.33\textwidth,clip=true]{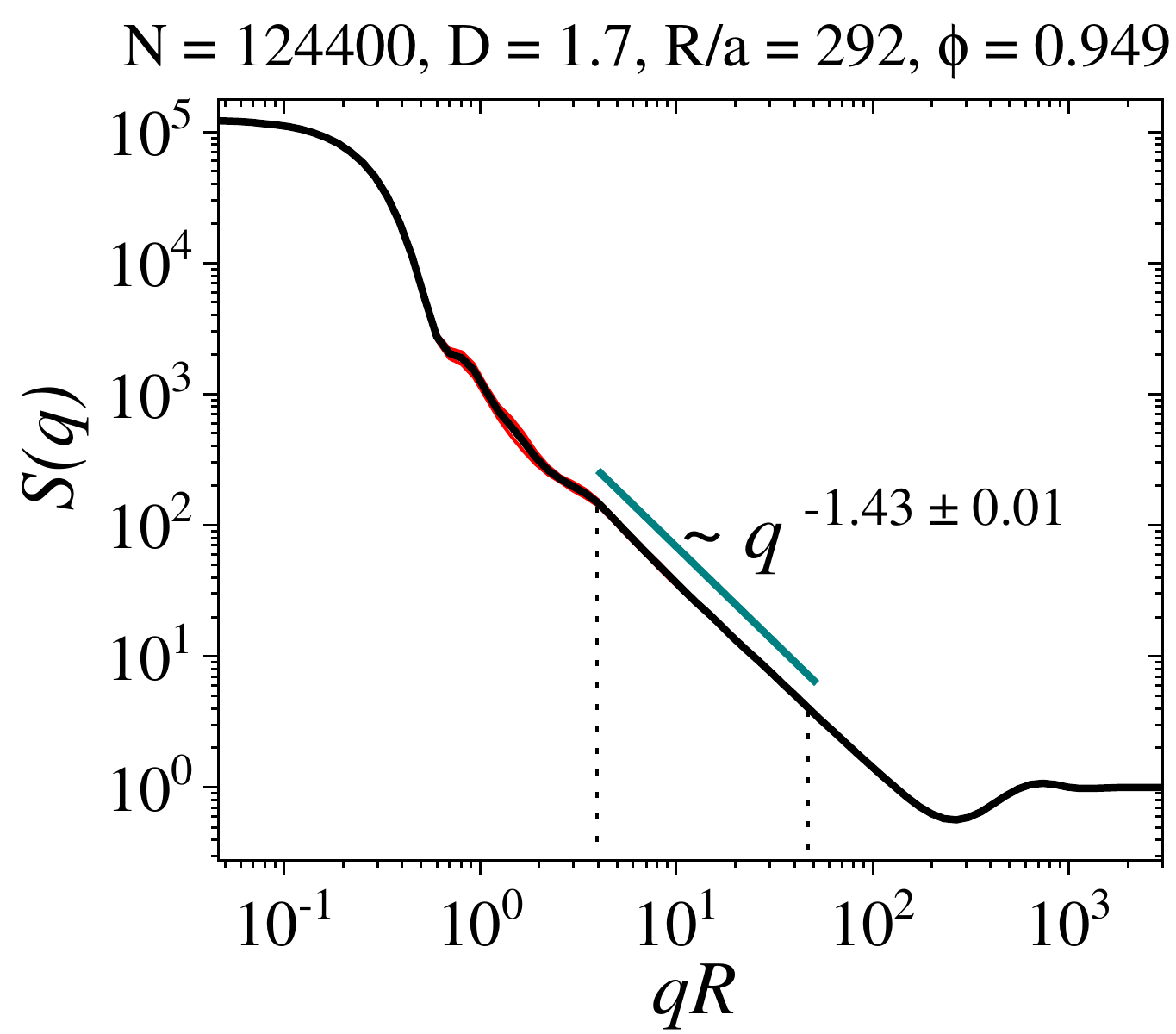} &
\includegraphics[width=0.33\textwidth,clip=true]{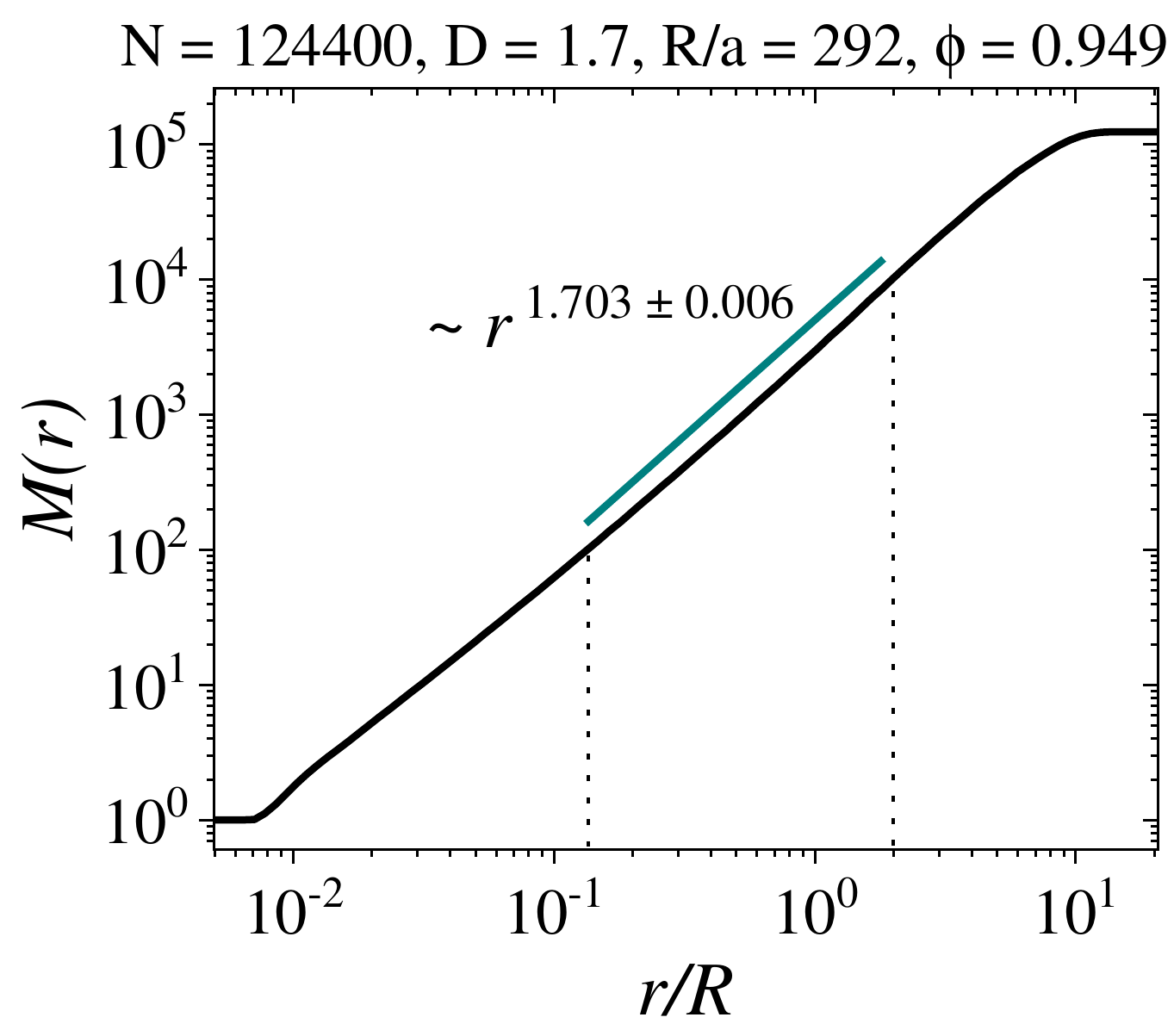} \\[4mm]

\includegraphics[width=0.28\textwidth,clip=true]{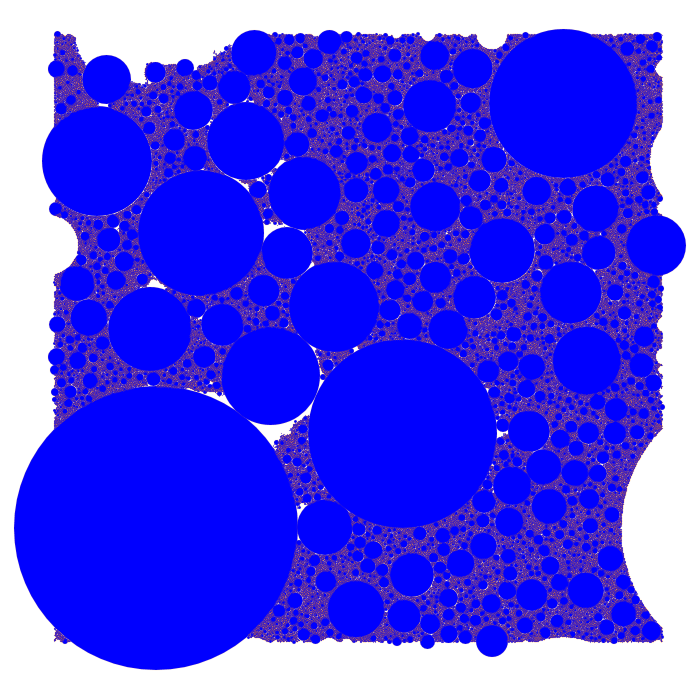} &
\includegraphics[width=0.33\textwidth,clip=true]{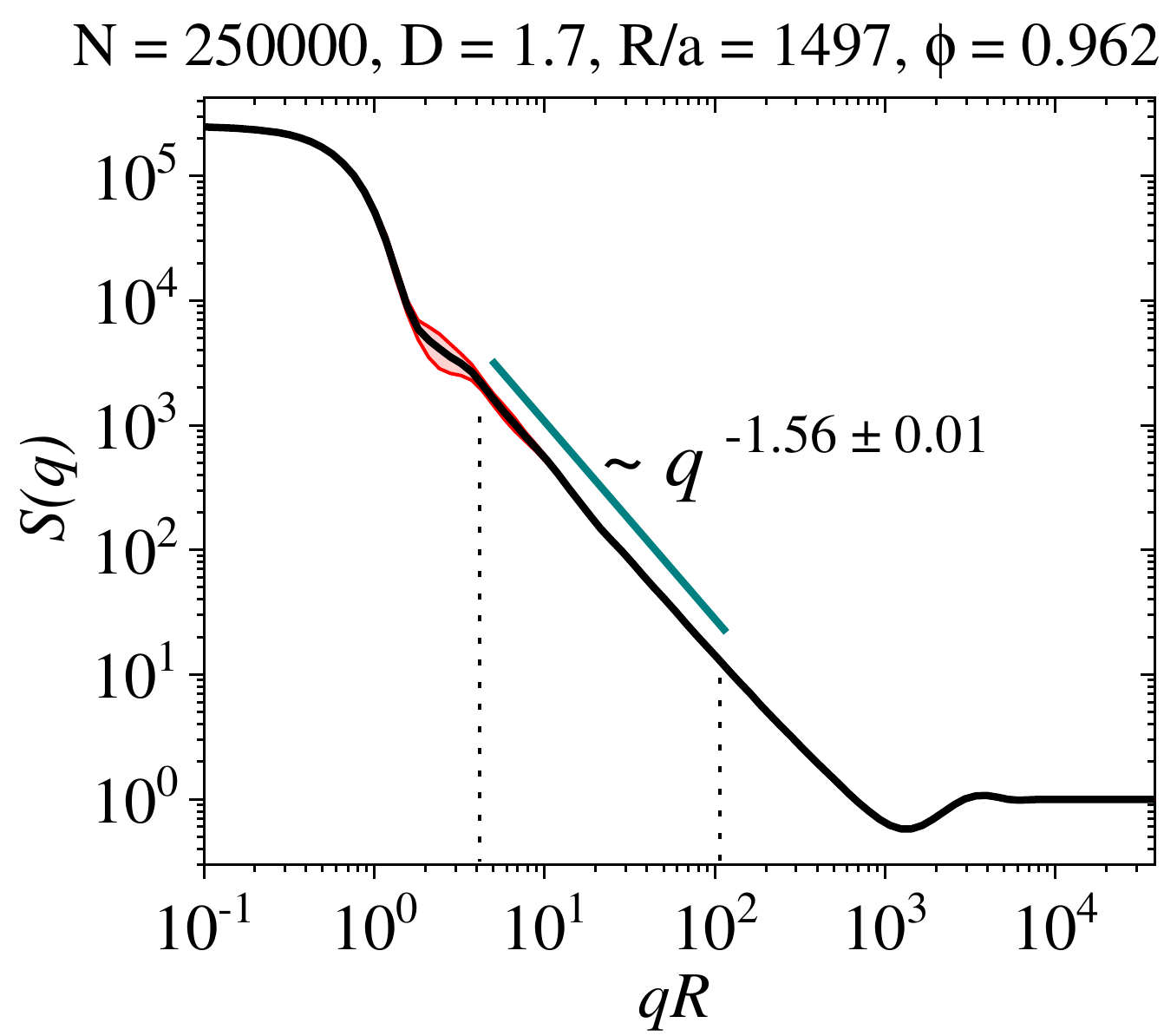} &
\includegraphics[width=0.34\textwidth,clip=true]{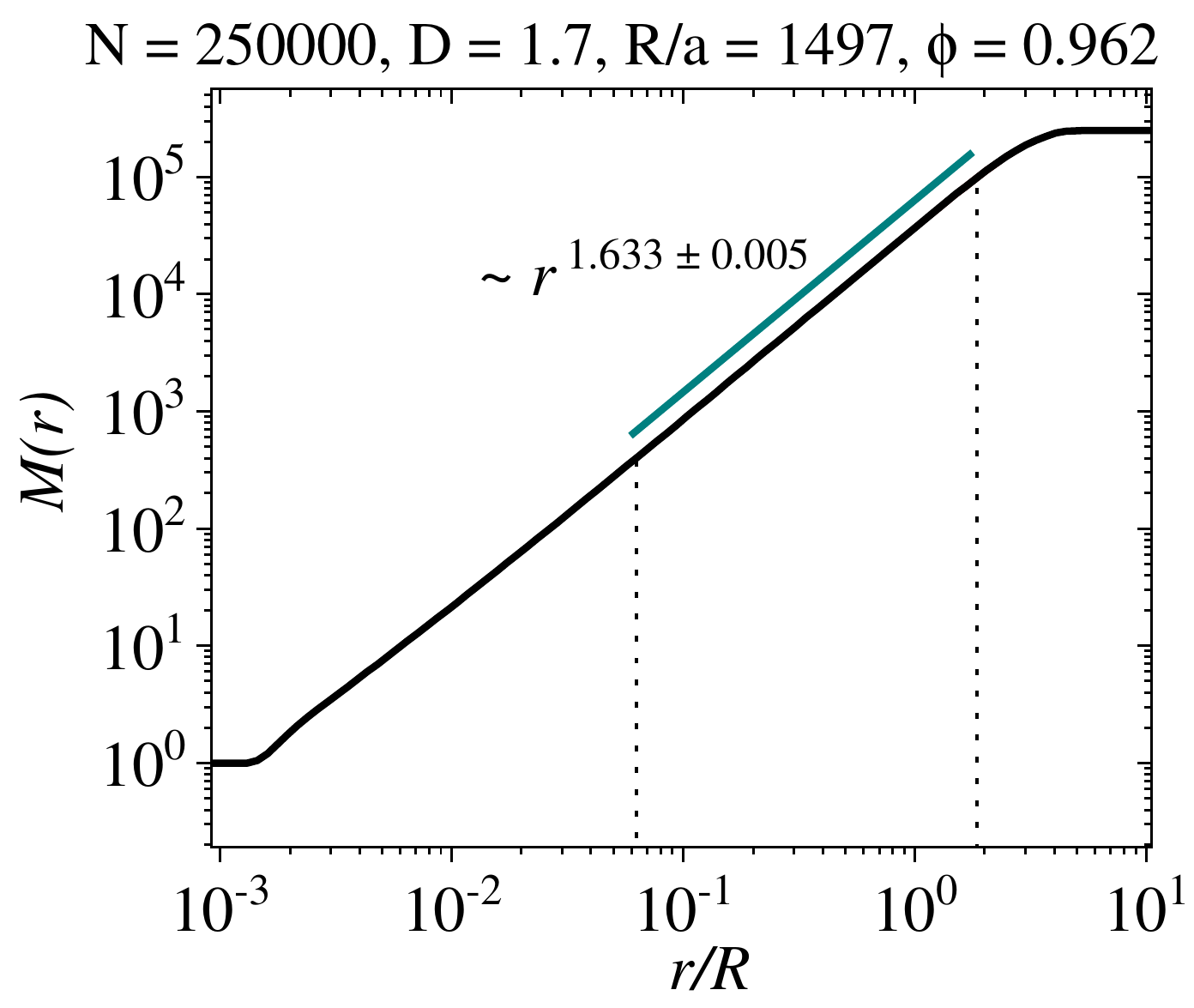}
\end{tabular}
\caption{\label{fig:S1}
Packing configurations for \(D=1.7\). Left column: spatial distribution of disks. Middle column: structure factor \(S(q)\). Right column: mass-radius relation \(M(r)\). Upper row: DT protocol with \(N=483000\). Middle row: CP protocol with \(N=124400\). Bottom row: CP protocol with \(N=250000\).
}
\end{figure}

\subsection{Purpose of the supplementary material}

This supplementary material provide additional packing configurations, the corresponding structure factors \(S(q)\) as functions of \(qR\), and the mass-radius relations \(M(r)\) as functions of \(r/R\). The results complement the discussion in the main text and Table \ref{tab:exponents} by showing how the Delaunay triangulation (DT) and constant-pressure (CP) protocols behave for different values of the distribution exponent \(D\). Besides, we compare the pores-size distribution and the combined pores and disks-size distribution for different size ratio in the CP protocol with $D=1.5$ (see Tables \ref{tab:exponents} and \ref{tab:pores} in the main text).

\begin{figure}[htbp]
\centering
\begin{tabular}{ccc}
\includegraphics[width=0.28\textwidth,clip=true]{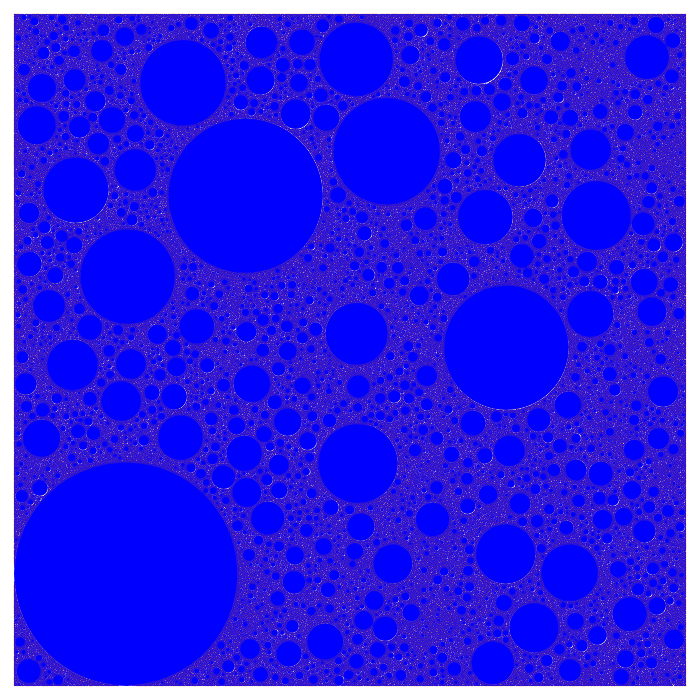} &
\includegraphics[width=0.33\textwidth,clip=true]{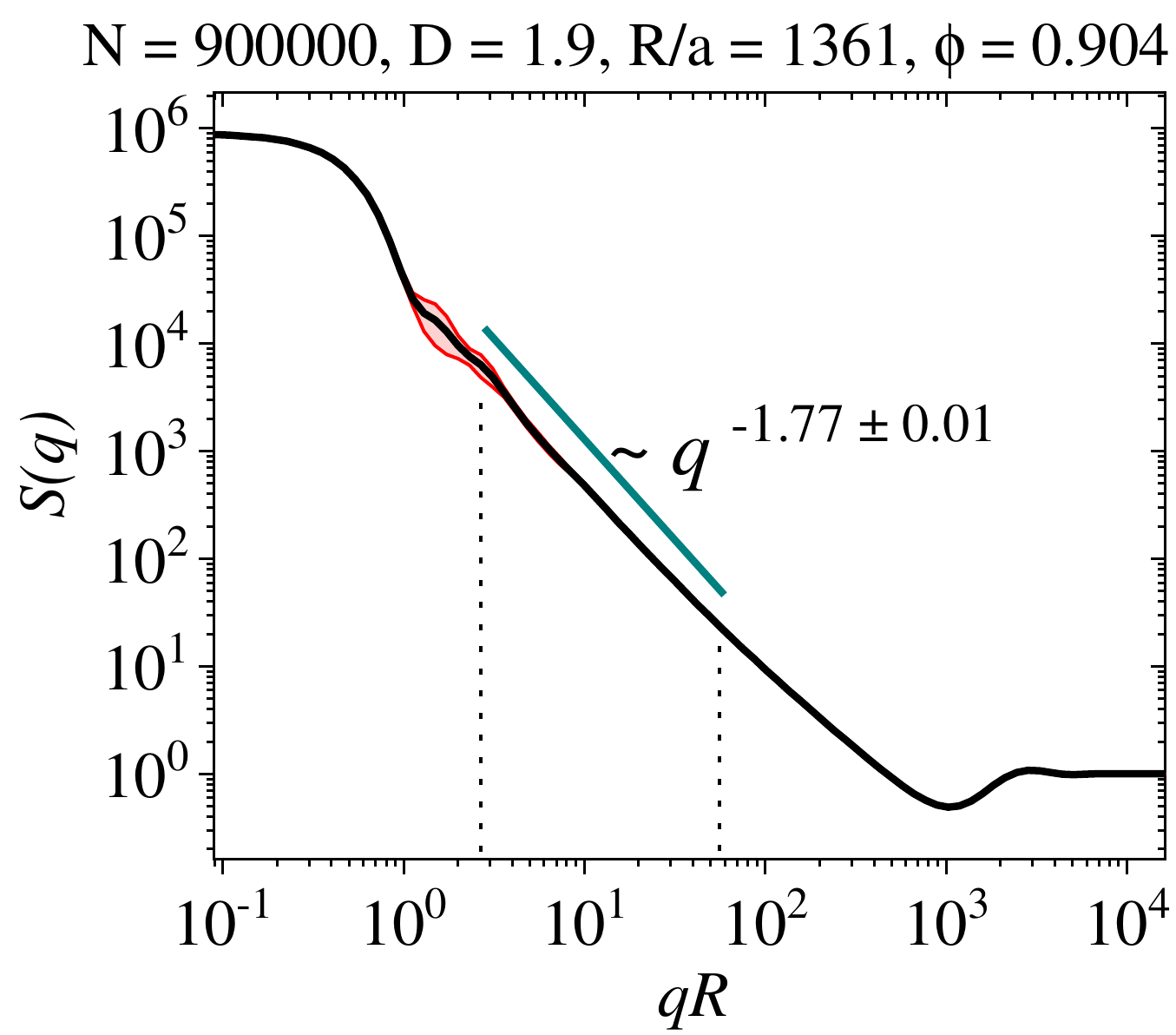} &
\includegraphics[width=0.33\textwidth,clip=true]{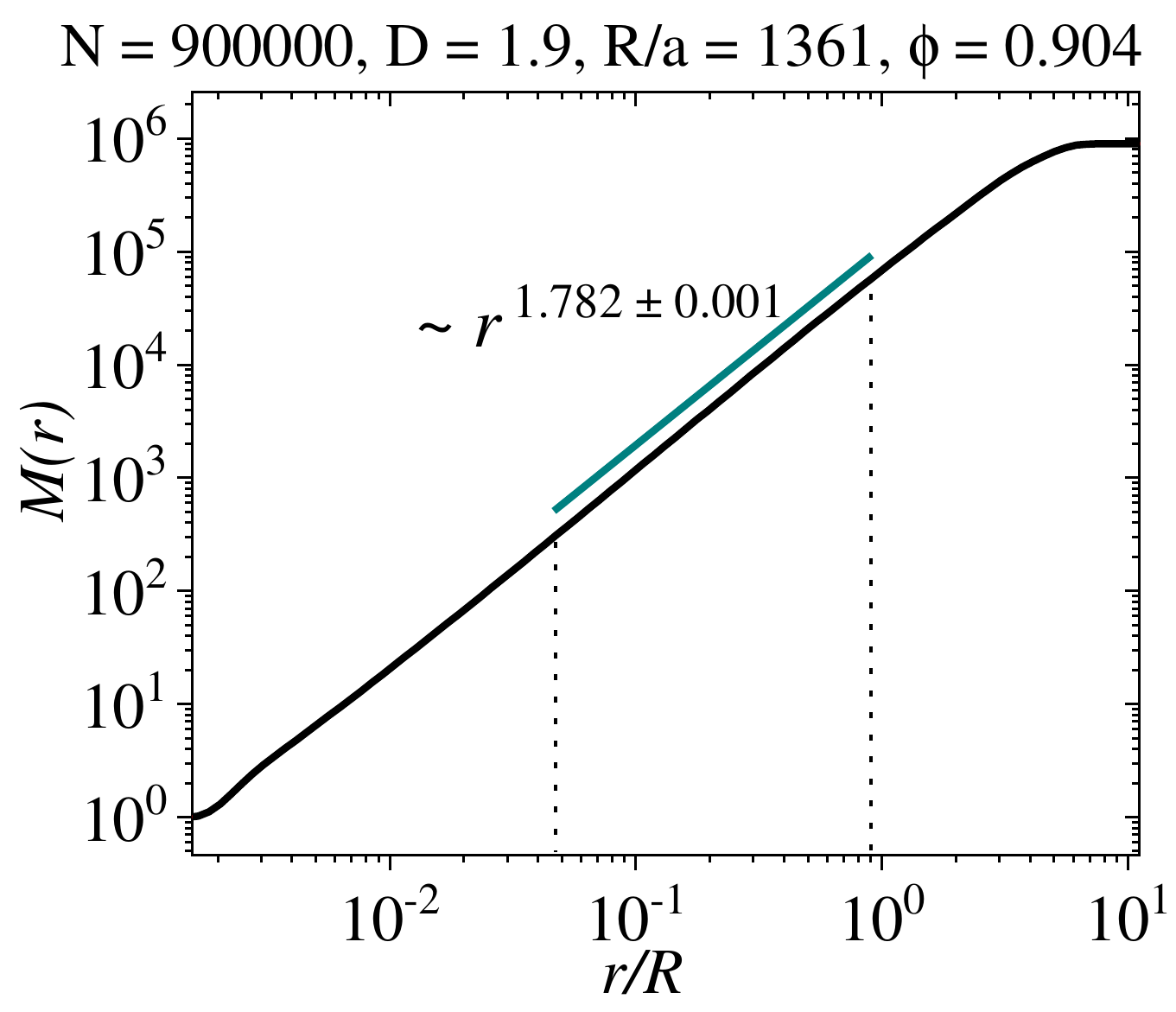} \\[4mm]

\includegraphics[width=0.28\textwidth,clip=true]{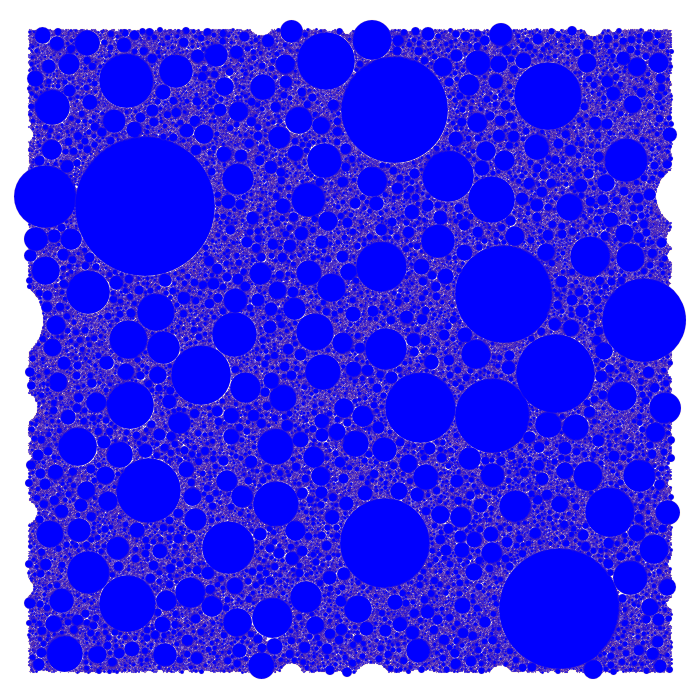} &
\includegraphics[width=0.33\textwidth,clip=true]{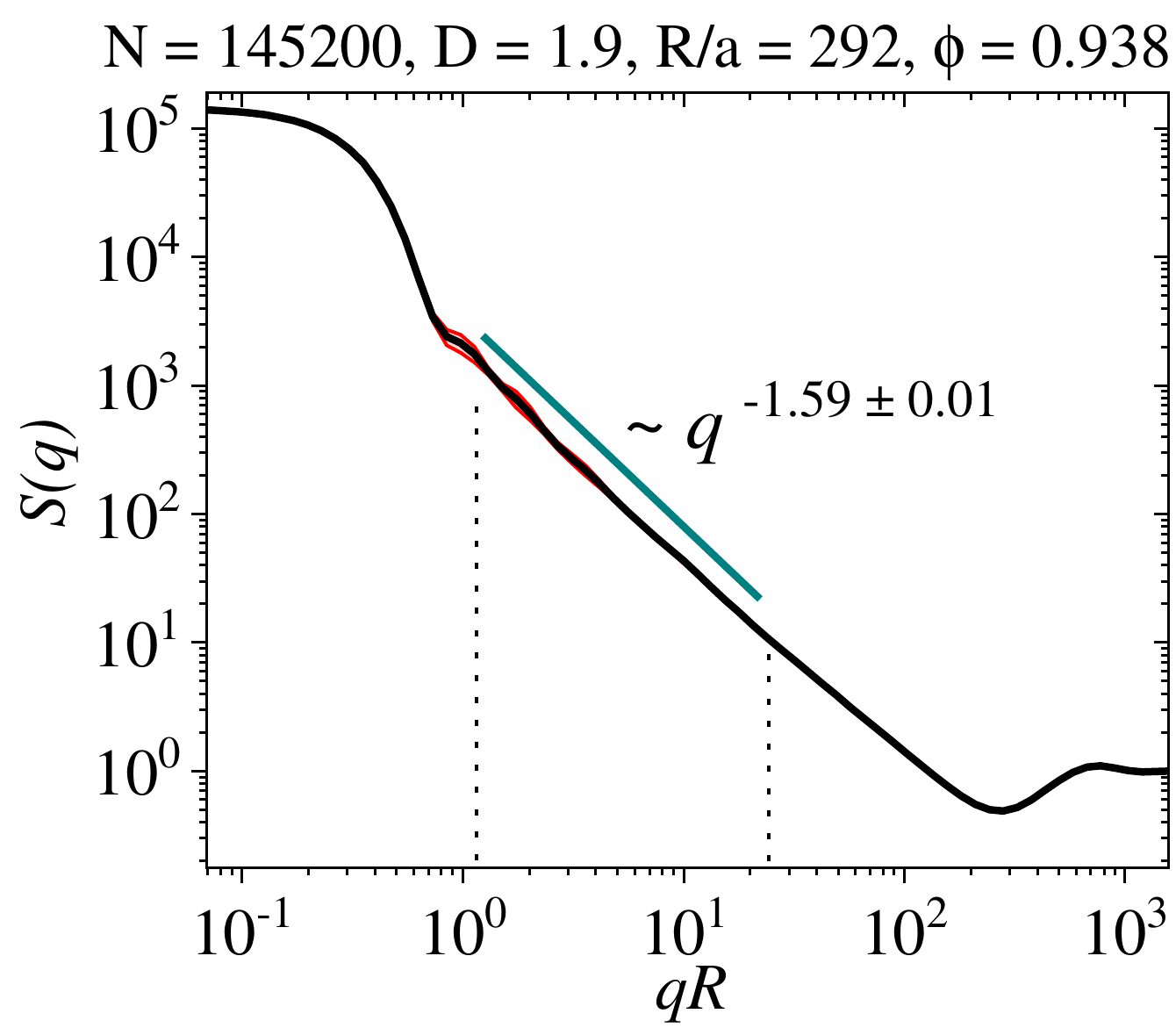} &
\includegraphics[width=0.33\textwidth,clip=true]{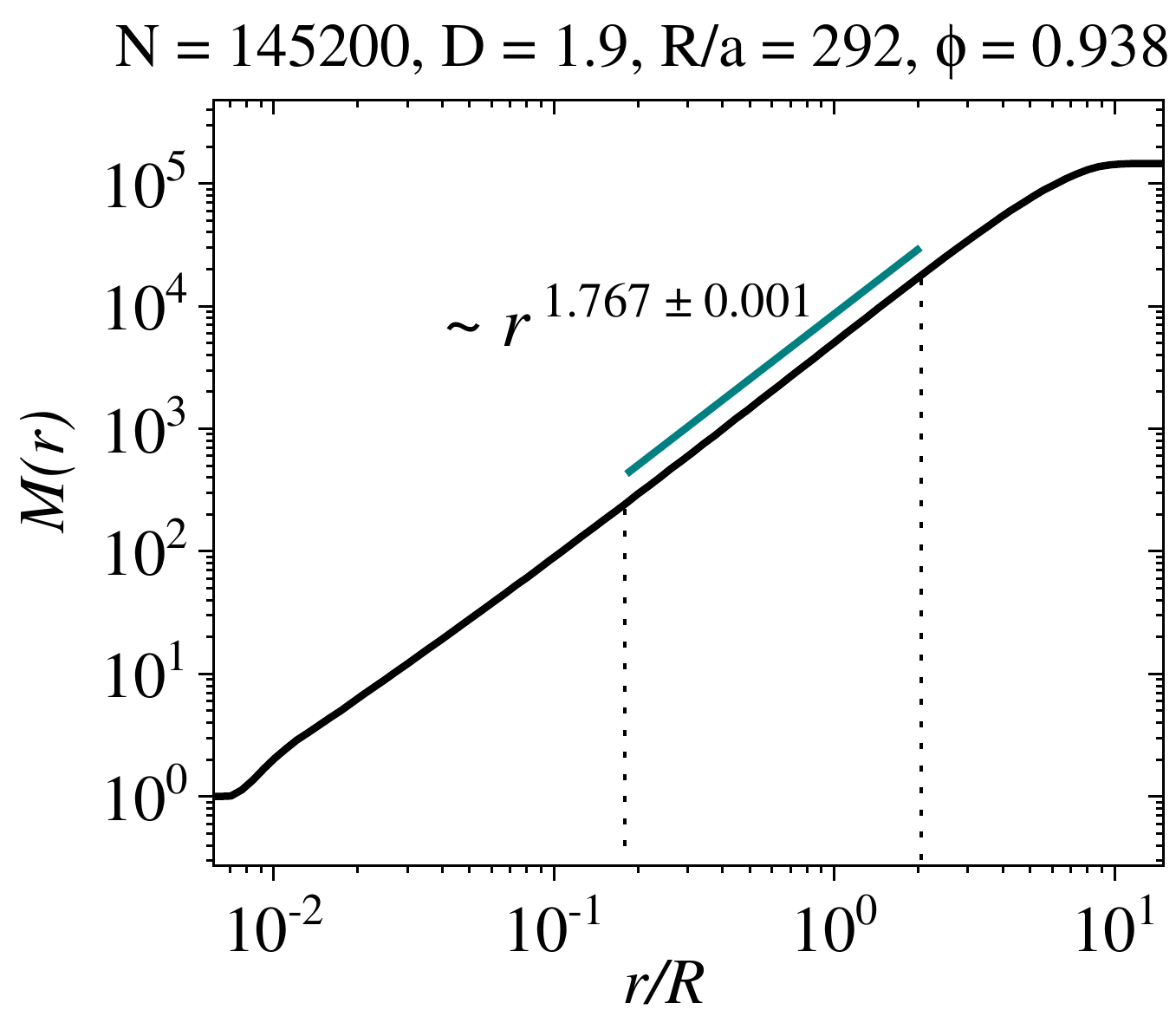} \\[4mm]

\includegraphics[width=0.28\textwidth,clip=true]{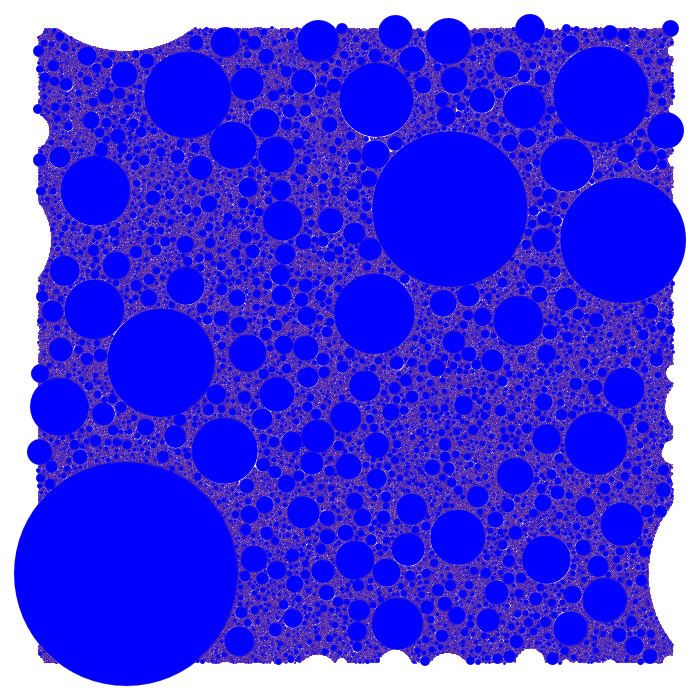} &
\includegraphics[width=0.33\textwidth,clip=true]{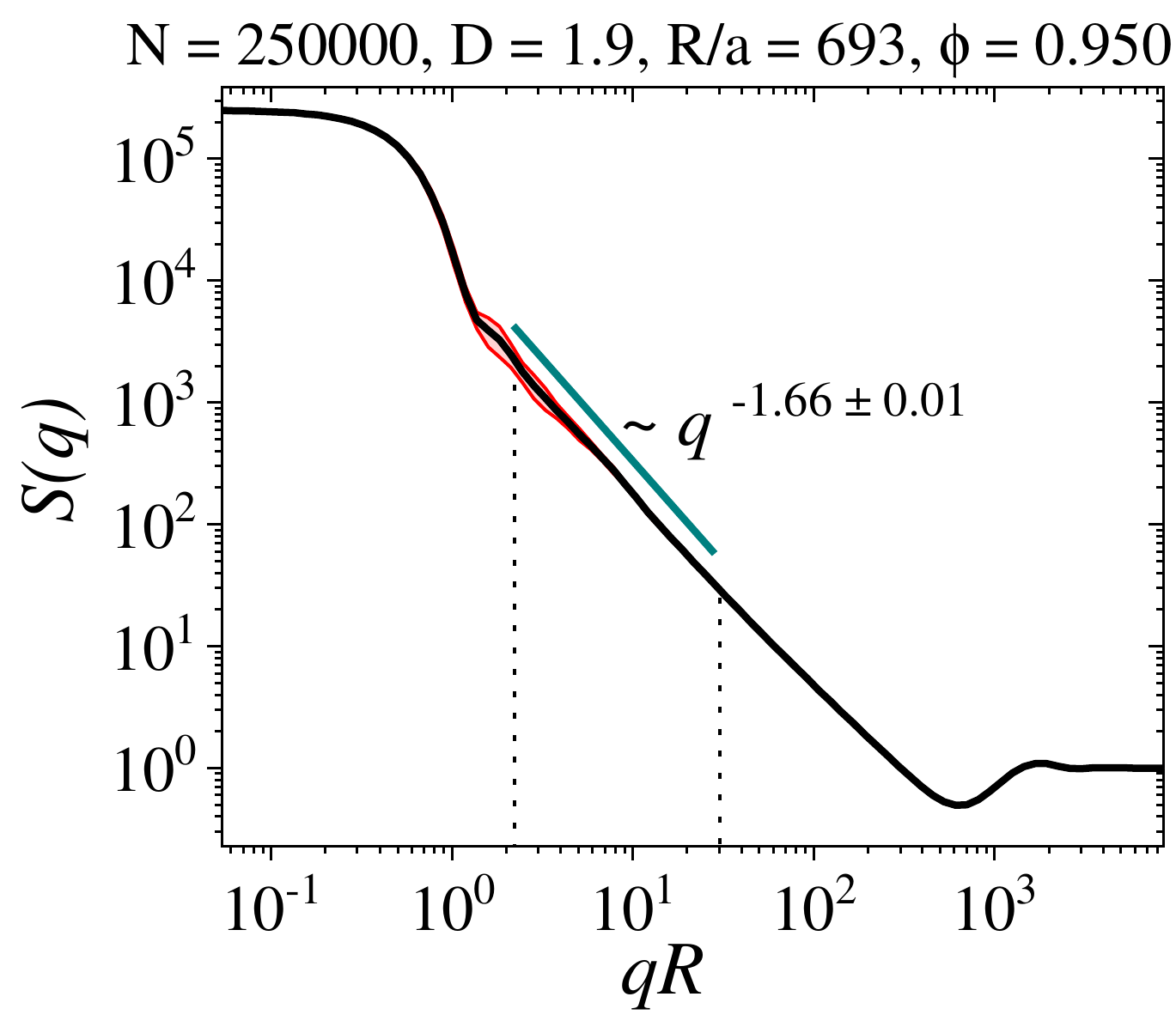} &
\includegraphics[width=0.34\textwidth,clip=true]{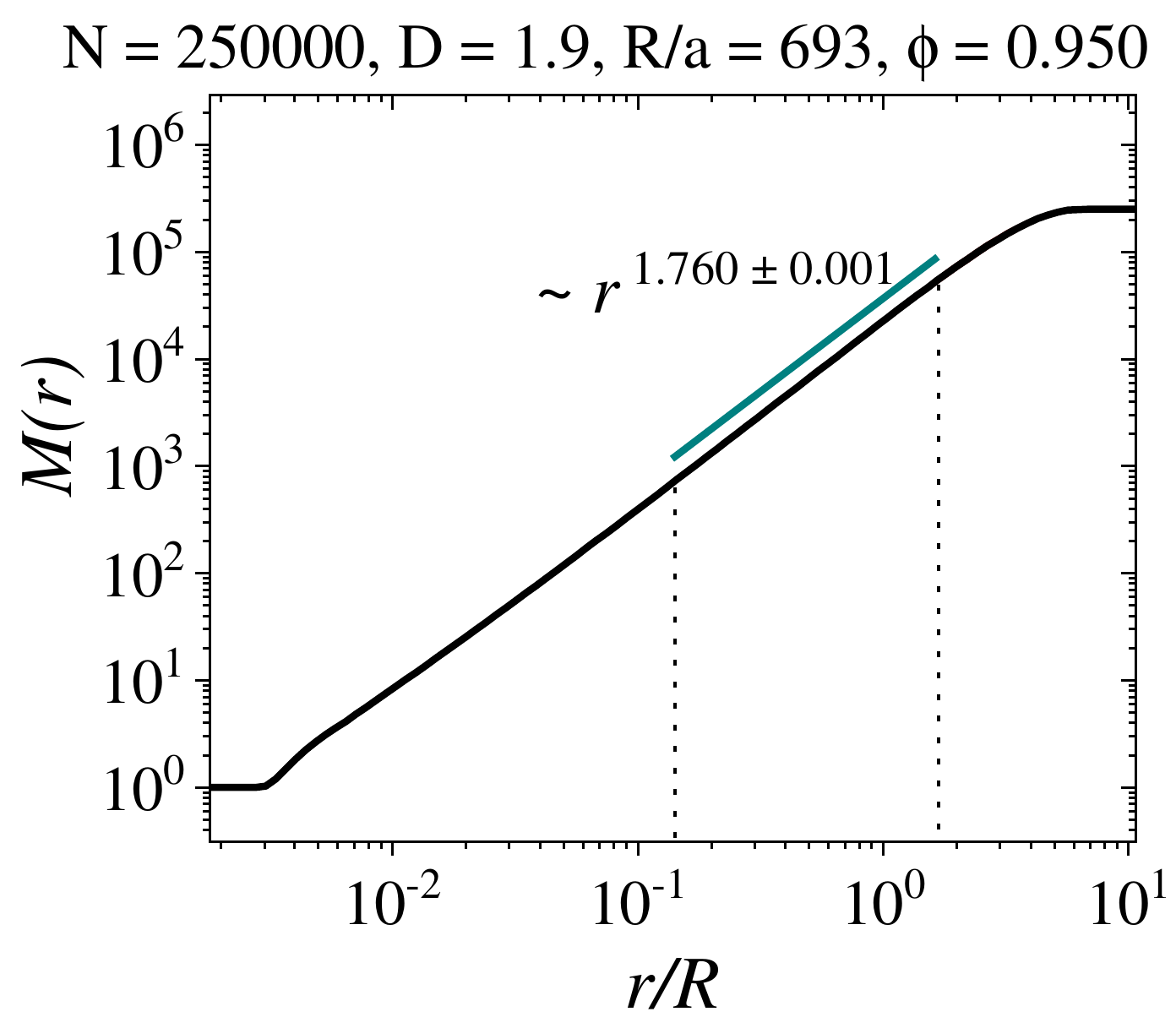}
\end{tabular}
\caption{\label{fig:S2}
Packing configurations for \(D=1.9\). Left column: spatial distribution of disks. Middle column: structure factor \(S(q)\). Right column: mass-radius relation \(M(r)\). Upper row: DT protocol with \(N=900000\). Middle row: CP protocol with \(N=145200\). Bottom row: CP protocol with \(N=250000\).
}
\end{figure}

\subsection{\(D=1.7\)}

Figure~\ref{fig:S1} shows additional packing configurations and the corresponding scaling functions for \(D=1.7\). The upper row corresponds to the DT protocol, while the middle and bottom rows correspond to CP packings with different numbers of disks.

\subsection{\(D=1.9\)}

Figure~\ref{fig:S2} shows the corresponding results for \(D=1.9\). As in Fig.~\ref{fig:S1}, the supplementary data include the spatial disk configurations, the structure factors, and the mass-radius relations for both DT and CP packings.

\subsection{Pore size distributions for the CP protocol with $D=1.5$}

\begin{figure}[h]
\centerline{\includegraphics[width=.6\columnwidth,clip=true]{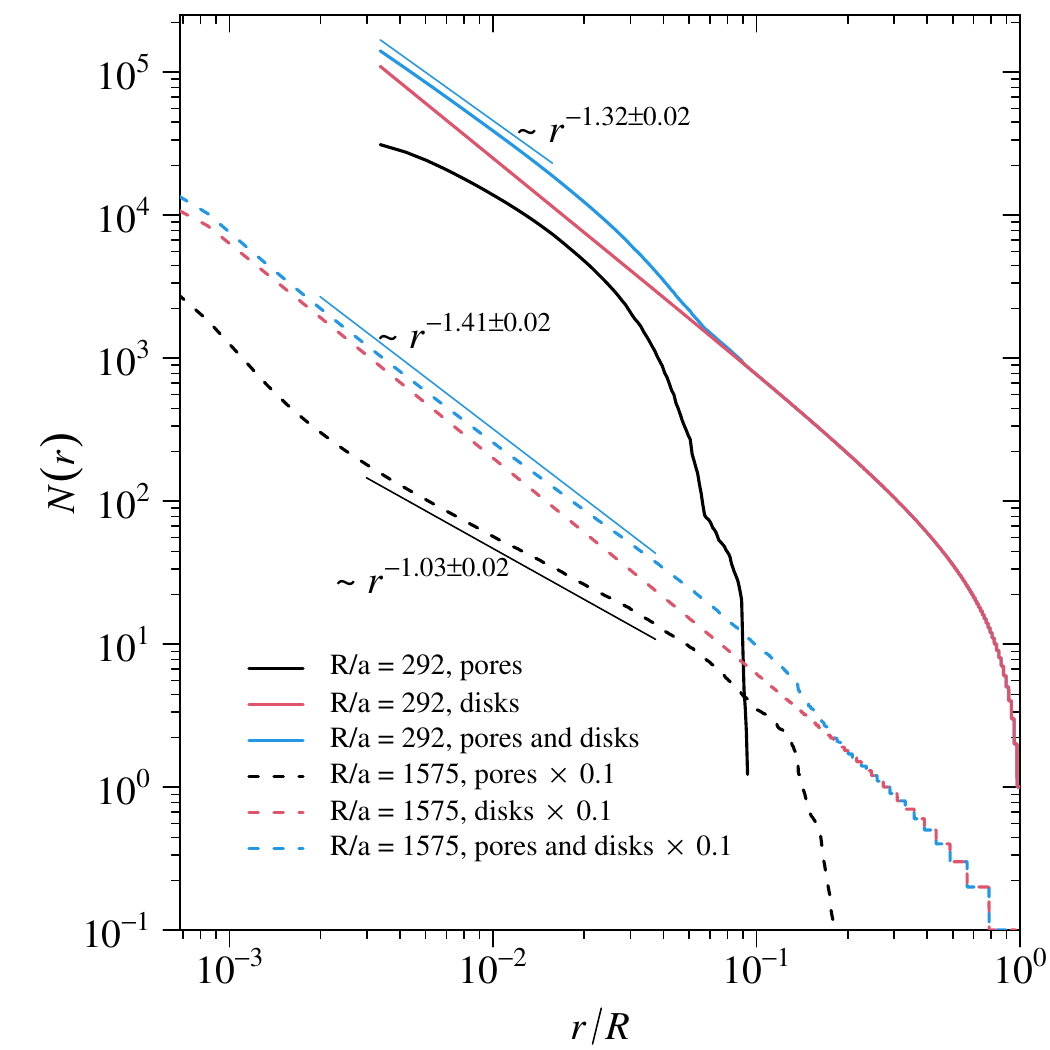}}
\caption{\label{fig:pores_comp}
Comparison of the pores-size distribution and the combined pores and disks-size distribution for different size ratio in the CP protocol with $D=1.5$. (see Sec.~VI in the main text for details). For the CP protocol with the ratio $R/a=292$, the fractal range for the pore size distribution $N_\mathrm{p}(r)$ practically shrinks to zero. Nevertheless, the combined pores and disks $N(r) + N_\mathrm{p}(r)$ size distribution contains a fractal range with the exponent $D_\mathrm{c}=1.32$. The decrease of the exponent from $D=1.5$ to $D_\mathrm{c}=1.32$ is due to the contribution of $N_\mathrm{p}(r)$. The curves are shifted vertically by a factor of 10 for better visualization.
}
\end{figure}

Figure~\ref{fig:pores_comp} compares the pores-size distribution and the combined pores and disks-size distribution for different size ratio in the CP protocol with D = 1.5.

\end{document}